\pdfoutput=1
\documentclass[12pt,a4paper]{article}

\usepackage{ifthen} 
\newboolean{pdflatex}
\setboolean{pdflatex}{true} 

\newboolean{articletitles}
\setboolean{articletitles}{true} 

\newboolean{uprightparticles}
\setboolean{uprightparticles}{false} 

\def\paperauthors{LHCb collaboration} 
\def\paperasciititle{Measurement of CP asymmetries and branching fraction ratios of Bm decays to two charm mesons} 
\def\papertitle{Measurement of \CP asymmetries and branching fraction ratios of \Bm decays to two charm mesons} 
\def\paperkeywords{{High Energy Physics}, {LHCb}} 
\def\papercopyright{\the\year\ CERN for the benefit of the LHCb collaboration} 
\def\paperlicence{CC BY 4.0 licence}
\def\paperlicenceurl{https://creativecommons.org/licenses/by/4.0/}

\usepackage[top=1in, bottom=1.25in, left=1in, right=1in]{geometry}

\usepackage{microtype}
\usepackage{lineno}  
\usepackage{xspace} 
\usepackage{caption} 

\usepackage{multirow}
\usepackage{array}

\usepackage{graphicx}  
\usepackage{color}
\usepackage{colortbl}
\graphicspath{{./figs/}} 

\usepackage{amsmath} 
\usepackage{amssymb}
\usepackage{amsfonts}
\usepackage{upgreek} 

\newcommand*\patchAmsMathEnvironmentForLineno[1]{%
\expandafter\let\csname old#1\expandafter\endcsname\csname #1\endcsname
\expandafter\let\csname oldend#1\expandafter\endcsname\csname
end#1\endcsname
 \renewenvironment{#1}%
   {\linenomath\csname old#1\endcsname}%
   {\csname oldend#1\endcsname\endlinenomath}%
}
\newcommand*\patchBothAmsMathEnvironmentsForLineno[1]{%
  \patchAmsMathEnvironmentForLineno{#1}%
  \patchAmsMathEnvironmentForLineno{#1*}%
}
\AtBeginDocument{%
\patchBothAmsMathEnvironmentsForLineno{equation}%
\patchBothAmsMathEnvironmentsForLineno{align}%
\patchBothAmsMathEnvironmentsForLineno{flalign}%
\patchBothAmsMathEnvironmentsForLineno{alignat}%
\patchBothAmsMathEnvironmentsForLineno{gather}%
\patchBothAmsMathEnvironmentsForLineno{multline}%
\patchBothAmsMathEnvironmentsForLineno{eqnarray}%
}

\usepackage{hyperxmp}

\usepackage[pdftex,
            pdfauthor={\paperauthors},
            pdftitle={\paperasciititle},
            pdfkeywords={\paperkeywords},
            pdfcopyright={Copyright (C) \papercopyright},
            pdflicenseurl={\paperlicenceurl}]{hyperref}

\usepackage[colorinlistoftodos,textsize=scriptsize]{todonotes}

\usepackage[bottom,flushmargin,hang,multiple]{footmisc}

\usepackage[all]{hypcap} 

\usepackage{xspace} 
\usepackage{upgreek}

\def\lhcb   {\mbox{LHCb}\xspace}

\def\babar  {\mbox{BaBar}\xspace}
\def\belle  {\mbox{Belle}\xspace}

\def\MagUp {\mbox{\em Mag\kern -0.05em Up}\xspace}

\ifthenelse{\boolean{uprightparticles}}%
{

 \def\Pmu         {\ensuremath{\upmu}\xspace}                 
 \def\Pnu         {\ensuremath{\upnu}\xspace}                 
                  
 \def\Ppi         {\ensuremath{\uppi}\xspace}

 \def\Ppsi        {\ensuremath{\uppsi}\xspace}

 \def\PDelta      {\ensuremath{\Delta}\xspace}                 
 \def\PXi         {\ensuremath{\Xi}\xspace}                 
 \def\PLambda     {\ensuremath{\Lambda}\xspace}                 
 \def\PSigma      {\ensuremath{\Sigma}\xspace}                 
 \def\POmega      {\ensuremath{\Omega}\xspace}                 
 \def\PUpsilon    {\ensuremath{\Upsilon}\xspace}
 \let\oldPi\Pi
 \def\PPi         {\ensuremath{\oldPi}\xspace}

 \def\PB      {\ensuremath{\mathrm{B}}\xspace}                 
                  
 \def\PD      {\ensuremath{\mathrm{D}}\xspace}

 \def\PJ      {\ensuremath{\mathrm{J}}\xspace}                 
 \def\PK      {\ensuremath{\mathrm{K}}\xspace}

 \def\PX      {\ensuremath{\mathrm{X}}\xspace}

 \def\Pb      {\ensuremath{\mathrm{b}}\xspace}                 
 \def\Pc      {\ensuremath{\mathrm{c}}\xspace}                 
 \def\Pd      {\ensuremath{\mathrm{d}}\xspace}

 \def\Pi      {\ensuremath{\mathrm{i}}\xspace}

 \def\Ps      {\ensuremath{\mathrm{s}}\xspace}

 \def\thebaroffset{0.0em}
}
{

 \def\Pmu         {\ensuremath{\mu}\xspace}                 
 \def\Pnu         {\ensuremath{\nu}\xspace}                 
                  
 \def\Ppi         {\ensuremath{\pi}\xspace}

 \def\Ppsi        {\ensuremath{\psi}\xspace}                 
                  
 \mathchardef\PDelta="7101
 \mathchardef\PXi="7104
 \mathchardef\PLambda="7103
 \mathchardef\PSigma="7106
 \mathchardef\POmega="710A
 \mathchardef\PUpsilon="7107
 \mathchardef\PPi="7105
                  
 \def\PB      {\ensuremath{B}\xspace}                 
                  
 \def\PD      {\ensuremath{D}\xspace}

 \def\PJ      {\ensuremath{J}\xspace}                 
 \def\PK      {\ensuremath{K}\xspace}

 \def\PX      {\ensuremath{X}\xspace}

 \def\Pb      {\ensuremath{b}\xspace}                 
 \def\Pc      {\ensuremath{c}\xspace}                 
 \def\Pd      {\ensuremath{d}\xspace}

 \def\Pi      {\ensuremath{i}\xspace}

 \def\Ps      {\ensuremath{s}\xspace}

 \def\thebaroffset{0.18em}
}
\newcommand{\offsetoverline}[2][\thebaroffset]{\kern #1\overline{\kern -#1 #2}}%

\makeatletter
\ifcase \@ptsize \relax
  \newcommand{\miniscule}{\@setfontsize\miniscule{4}{5}}
\or
  \newcommand{\miniscule}{\@setfontsize\miniscule{5}{6}}
\or
  \newcommand{\miniscule}{\@setfontsize\miniscule{5}{6}}
\fi
\makeatother

\DeclareRobustCommand{\optbar}[1]{\shortstack{{\miniscule (\rule[.5ex]{1.25em}{.18mm})}
  \\ [-.7ex] $#1$}}

\def\mup        {{\ensuremath{\Pmu^+}}\xspace}
\def\mun        {{\ensuremath{\Pmu^-}}\xspace} 

\def\neu        {{\ensuremath{\Pnu}}\xspace}

\def\neum       {{\ensuremath{\neu_\mu}}\xspace}

\def\dquark    {{\ensuremath{\Pd}}\xspace}

\def\squark    {{\ensuremath{\Ps}}\xspace}

\def\cquark    {{\ensuremath{\Pc}}\xspace}
\def\cquarkbar {{\ensuremath{\overline \cquark}}\xspace}

\def\bquark    {{\ensuremath{\Pb}}\xspace}

\def\pion   {{\ensuremath{\Ppi}}\xspace}
\def\piz    {{\ensuremath{\pion^0}}\xspace}
\def\pip    {{\ensuremath{\pion^+}}\xspace}
\def\pim    {{\ensuremath{\pion^-}}\xspace}

\def\kaon    {{\ensuremath{\PK}}\xspace}
\def\Kbar    {{\ensuremath{\offsetoverline{\PK}}}\xspace}

\def\KorKbar {\kern \thebaroffset\optbar{\kern -\thebaroffset \PK}{}\xspace}

\def\Kzb     {{\ensuremath{\Kbar{}^0}}\xspace}
\def\Kp      {{\ensuremath{\kaon^+}}\xspace}
\def\Km      {{\ensuremath{\kaon^-}}\xspace}

\def\Dbar    {{\ensuremath{\offsetoverline{\PD}}}\xspace}
\def\D       {{\ensuremath{\PD}}\xspace}
\def\Db      {{\ensuremath{\Dbar}}\xspace}
\def\DorDbar {\kern \thebaroffset\optbar{\kern -\thebaroffset \PD}\xspace}
\def\Dz      {{\ensuremath{\D^0}}\xspace}
\def\Dzb     {{\ensuremath{\Dbar{}^0}}\xspace}
\def\Dp      {{\ensuremath{\D^+}}\xspace}
\def\Dm      {{\ensuremath{\D^-}}\xspace}

\def\DpDm    {\ensuremath{\Dp {\kern -0.16em \Dm}}\xspace}
\def\Dstar   {{\ensuremath{\D^*}}\xspace}

\def\Dstarz  {{\ensuremath{\D^{*0}}}\xspace}

\def\Dstarp  {{\ensuremath{\D^{*+}}}\xspace}
\def\Dstarm  {{\ensuremath{\D^{*-}}}\xspace}
\def\Dstarpm {{\ensuremath{\D^{*\pm}}}\xspace}

\def\Dsm     {{\ensuremath{\D^-_\squark}}\xspace}

\def\Dssm    {{\ensuremath{\D^{*-}_\squark}}\xspace}

\def\B       {{\ensuremath{\PB}}\xspace}

\def\BorBbar {\kern \thebaroffset\optbar{\kern -\thebaroffset \PB}\xspace}
\def\Bz      {{\ensuremath{\B^0}}\xspace}

\def\Bd      {{\ensuremath{\B^0}}\xspace}

\def\BdorBdbar {\kern \thebaroffset\optbar{\kern -\thebaroffset \Bd}\xspace}
\def\Bu      {{\ensuremath{\B^+}}\xspace}
\def\Bub     {{\ensuremath{\B^-}}\xspace}
\def\Bp      {{\ensuremath{\Bu}}\xspace}
\def\Bm      {{\ensuremath{\Bub}}\xspace}
\def\Bpm     {{\ensuremath{\B^\pm}}\xspace}

\def\Bs      {{\ensuremath{\B^0_\squark}}\xspace}

\def\BsorBsbar {\kern \thebaroffset\optbar{\kern -\thebaroffset \Bs}\xspace}

\def\BdorBs  {\Bds}

\def\jpsi     {{\ensuremath{{\PJ\mskip -3mu/\mskip -2mu\Ppsi}}}\xspace}

\def\Y#1S{\ensuremath{\PUpsilon{(#1S)}}\xspace}

\def\LorLbar     {\kern \thebaroffset\optbar{\kern -\thebaroffset \PLambda}\xspace}

\def\BF         {{\ensuremath{\mathcal{B}}}\xspace}

\newcommand{\decay}[2]{\ensuremath{#1\!\to #2}\xspace} 

\def\to                 {\ensuremath{\rightarrow}\xspace}

\def\CP                {{\ensuremath{C\!P}}\xspace}

\newcommand{\ACP}{{\ensuremath{{\mathcal{A}}^{\CP}}}\xspace}

\def\AT#1     {\ensuremath{A_{\mathrm{T}}^{#1}}\xspace}           

\def\C#1      {\ensuremath{\mathcal{C}_{#1}}\xspace}                       
\def\Cp#1     {\ensuremath{\mathcal{C}_{#1}^{'}}\xspace}                    
\def\Ceff#1   {\ensuremath{\mathcal{C}_{#1}^{\mathrm{(eff)}}}\xspace}        
\def\Cpeff#1  {\ensuremath{\mathcal{C}_{#1}^{'\mathrm{(eff)}}}\xspace}       
\def\Ope#1    {\ensuremath{\mathcal{O}_{#1}}\xspace}                       
\def\Opep#1   {\ensuremath{\mathcal{O}_{#1}^{'}}\xspace}                    

\newcommand{\nospaceunit}[1]{\ensuremath{\text{#1}}}       
\newcommand{\aunit}[1]{\ensuremath{\text{\,#1}}}       

\newcommand{\tev}{\aunit{Te\kern -0.1em V}\xspace}
\newcommand{\gev}{\aunit{Ge\kern -0.1em V}\xspace}
\newcommand{\mev}{\aunit{Me\kern -0.1em V}\xspace}
\newcommand{\kev}{\aunit{ke\kern -0.1em V}\xspace}
\newcommand{\ev}{\aunit{e\kern -0.1em V}\xspace}
 
\newcommand{\mevc}{\ensuremath{\aunit{Me\kern -0.1em V\!/}c}\xspace}
\newcommand{\gevc}{\ensuremath{\aunit{Ge\kern -0.1em V\!/}c}\xspace}
\newcommand{\mevcc}{\ensuremath{\aunit{Me\kern -0.1em V\!/}c^2}\xspace}
\newcommand{\gevcc}{\ensuremath{\aunit{Ge\kern -0.1em V\!/}c^2}\xspace}

\def\mum  {\ensuremath{\,\upmu\nospaceunit{m}}\xspace}

\def\fb   {\ensuremath{\aunit{fb}}\xspace}
\def\invfb   {\ensuremath{\fb^{-1}}\xspace}

\newcommand{\chisq}{\ensuremath{\chi^2}\xspace}

\newcommand{\chisqip}{\ensuremath{\chi^2_{\text{IP}}}\xspace}

\def\gsim{{~\raise.15em\hbox{$>$}\kern-.85em
          \lower.35em\hbox{$\sim$}~}\xspace}
\def\lsim{{~\raise.15em\hbox{$<$}\kern-.85em
          \lower.35em\hbox{$\sim$}~}\xspace}

\def\pt         {\ensuremath{p_{\mathrm{T}}}\xspace}

\def\ptot       {\ensuremath{p}\xspace}

\def\dllkpi     {\ensuremath{\mathrm{DLL}_{\kaon\pion}}\xspace}

\def\mrad{\aunit{mrad}\xspace}

\def\evtgen     {\mbox{\textsc{EvtGen}}\xspace}

\def\geant      {\mbox{\textsc{Geant4}}\xspace}

\def\photos     {\mbox{\textsc{Photos}}\xspace}

\def\pythia     {\mbox{\textsc{Pythia}}\xspace}

\def\tell1  {TELL1\xspace}
\def\ukl1   {UKL1\xspace}

\newcommand{\lhcborcid}[1]{\href{https://orcid.org/#1}{\hspace*{0.1em}\raisebox{-0.45ex}{\includegraphics[width=1em]{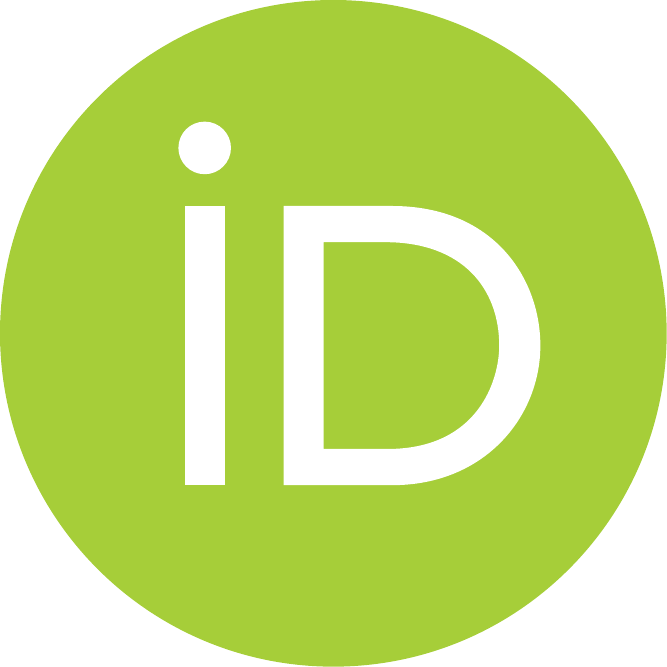}}}}

\usepackage{cite} 
\usepackage{mciteplus}

\def\DmorDsorDstar{{\ensuremath{\D^{(*)-}_{(\squark)}}}\xspace}
\def\DporDsorDstar{{\ensuremath{\D^{(*)+}_{(\squark)}}}\xspace}
\def\DmorDs{{\ensuremath{\D^-_{(\squark)}}}\xspace}
\def\DmorDstar{{\ensuremath{\D^{(*)-}}}\xspace}
\def\DsmorDstar{{\ensuremath{\D^{(*)-}_{\squark}}}\xspace}

\def\DzorDstar{{\ensuremath{\D^{(*)0}}}\xspace}
\def\DzborDstar{{\ensuremath{\Db^{(*)0}}}\xspace}
\def\DorDstar{{\ensuremath{\D^{(*)}}}\xspace}

\def\DmporDstar{{\ensuremath{\D^{(*)\mp}}}\xspace}
\def\DKpi{{\ensuremath{\mbox{\decay{\Dz}{K^-\pi^+}}}}\xspace}
\def\DKpipipi{{\ensuremath{\mbox{\decay{\Dz}{K^-\pi^+\pi^-\pi^+}}}}\xspace}
\def\Dstst{{\ensuremath{\D^{**}}}\xspace}
\def\BdorBs{{\ensuremath{\B^0_{(\squark)}}}\xspace}
\newcommand{\Araw}{{\ensuremath{{\mathcal{A}}_{\mathrm{ raw}}}}\xspace}
\newcommand{\AD}{{\ensuremath{{\mathcal{A}}_{\mathrm{ D}}}}\xspace}
\newcommand{\AKpi}{{\ensuremath{{\mathcal{A}}_{K\pi}}}\xspace}
\newcommand{\Api}{{\ensuremath{{\mathcal{A}}_{\pi}}}\xspace}
\newcommand{\AP}{{\ensuremath{{\mathcal{A}}_{\mathrm{ P}}}}\xspace}
\newcommand{\APID}{{\ensuremath{{\mathcal{A}}_{\mathrm{ PID}}}}\xspace}
\newcommand{\ATIS}{{\ensuremath{{\mathcal{A}}_{\mathrm{TIS}}}}\xspace}
\newcommand{\ATOS}{{\ensuremath{{\mathcal{A}}_{\mathrm{TOS}}}}\xspace}

\begin{document}

\renewcommand{\thefootnote}{\fnsymbol{footnote}}
\setcounter{footnote}{1}


\begin{titlepage}
\pagenumbering{roman}

\vspace*{-1.5cm}
\centerline{\large EUROPEAN ORGANIZATION FOR NUCLEAR RESEARCH (CERN)}
\vspace*{1.5cm}
\noindent
\begin{tabular*}{\linewidth}{lc@{\extracolsep{\fill}}r@{\extracolsep{0pt}}}
\ifthenelse{\boolean{pdflatex}}
{\vspace*{-1.5cm}\mbox{\!\!\!\includegraphics[width=.14\textwidth]{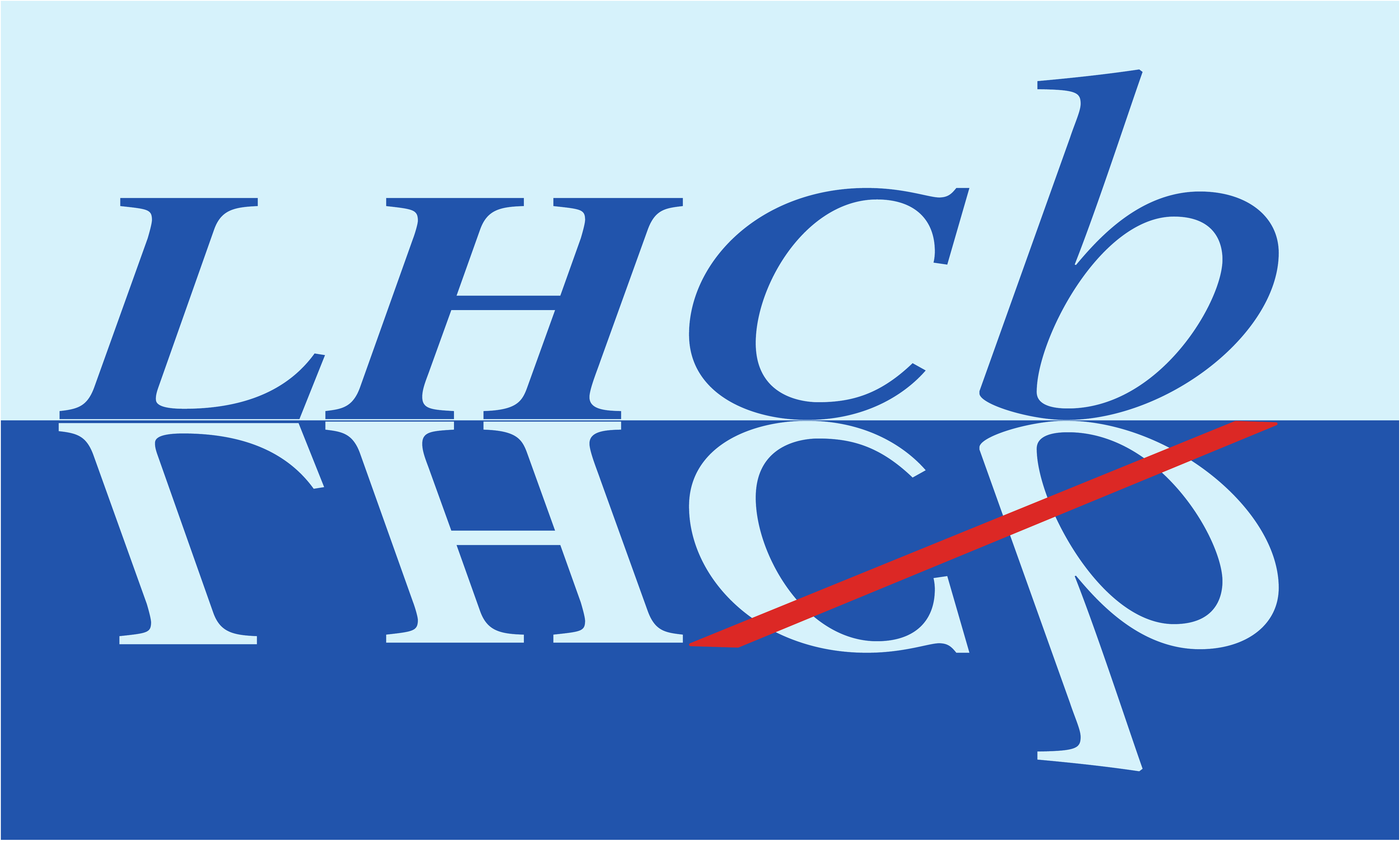}} & &}%
{\vspace*{-1.2cm}\mbox{\!\!\!\includegraphics[width=.12\textwidth]{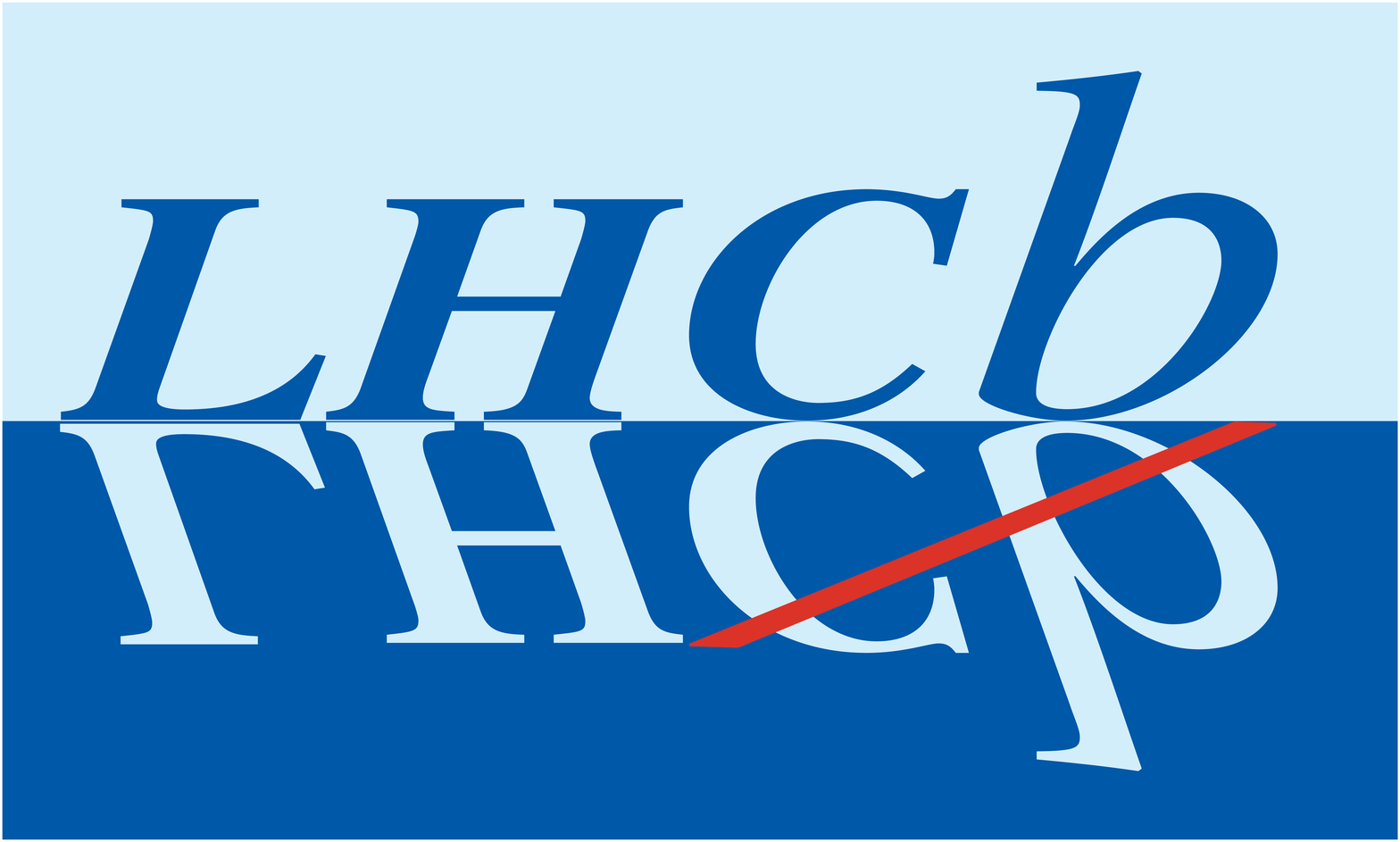}} & &}%
\\
 & & CERN-EP-2023-104 \\  
 & & LHCb-PAPER-2023-007 \\  
 & & September 28, 2023 \\ 
\end{tabular*}

\vspace*{4.0cm}

{\normalfont\bfseries\boldmath\huge
\begin{center}
  \papertitle 
\end{center}
}

\vspace*{2.0cm}

\begin{center}
\paperauthors\footnote{Authors are listed at the end of this paper.}
\end{center}

\vspace{\fill}

\begin{abstract}
  \noindent
The \CP asymmetries of seven 
\Bm decays to two charm mesons are measured using data corresponding to an integrated luminosity of \mbox{$9\invfb$} of proton-proton collisions collected by the \lhcb experiment.
Decays involving a \Dstarz or \Dssm meson are analysed by reconstructing only the \Dz or \Dsm decay products.
This paper presents the first measurement of \mbox{$\ACP(\decay{\Bm}{\Dssm\Dz})$} and \mbox{$\ACP(\decay{\Bm}{\Dsm\Dstarz})$}, and the most precise measurement of the other five \CP asymmetries.
There is no evidence of \CP violation in any of the analysed decays.
Additionally, two ratios between branching fractions of selected decays are measured.  
\end{abstract}

\vspace*{2.0cm}

\begin{center}
  Published in JHEP 09 (2023) 202
\end{center}

\vspace{\fill}

{\footnotesize 
\centerline{\copyright~\papercopyright. \href{\paperlicenceurl}{\paperlicence}.}}
\vspace*{2mm}

\end{titlepage}


\newpage
\setcounter{page}{2}
\mbox{~}
%
%
%
%


\renewcommand{\thefootnote}{\arabic{footnote}}
\setcounter{footnote}{0}

\cleardoublepage


\pagestyle{plain} 
\setcounter{page}{1}
\pagenumbering{arabic}



\section{Introduction}
\label{sec:Introduction}

The breaking of the combined charge-parity (\CP) symmetry in weak interactions arises in the Standard Model (SM) from a single irreducible complex phase in the Cabibbo-Kobayashi-Maskawa matrix~\cite{Cabibbo:1963yz,Kobayashi:1973fv}.
Direct \CP violation, namely differences in the decay rates between \CP-conjugate processes, results from the interference between two or more amplitudes that have different weak and strong phases.
The \CP asymmetries in \mbox{\decay{\Bm}{\DmorDsorDstar\DzorDstar}} decays\footnote{The inclusion of charge-conjugate processes is implied throughout except in the discussion of asymmetries.} are defined as
\begin{equation}
\ACP\equiv\frac{\Gamma(\decay{\Bm}{\DmorDsorDstar\DzorDstar})-\Gamma(\decay{\Bp}{\DporDsorDstar\DzborDstar})}{\Gamma(\decay{\Bm}{\DmorDsorDstar\DzorDstar})+\Gamma(\decay{\Bp}{\DporDsorDstar\DzborDstar})},
\label{eq:ACPdef}
\end{equation}
where \Dstarm refers to the $D^*(2010)^-$ and \Dstarz to the $D^*(2007)^0$ mesons.

\begin{figure}[bp]
\centering
\includegraphics[width=5cm]{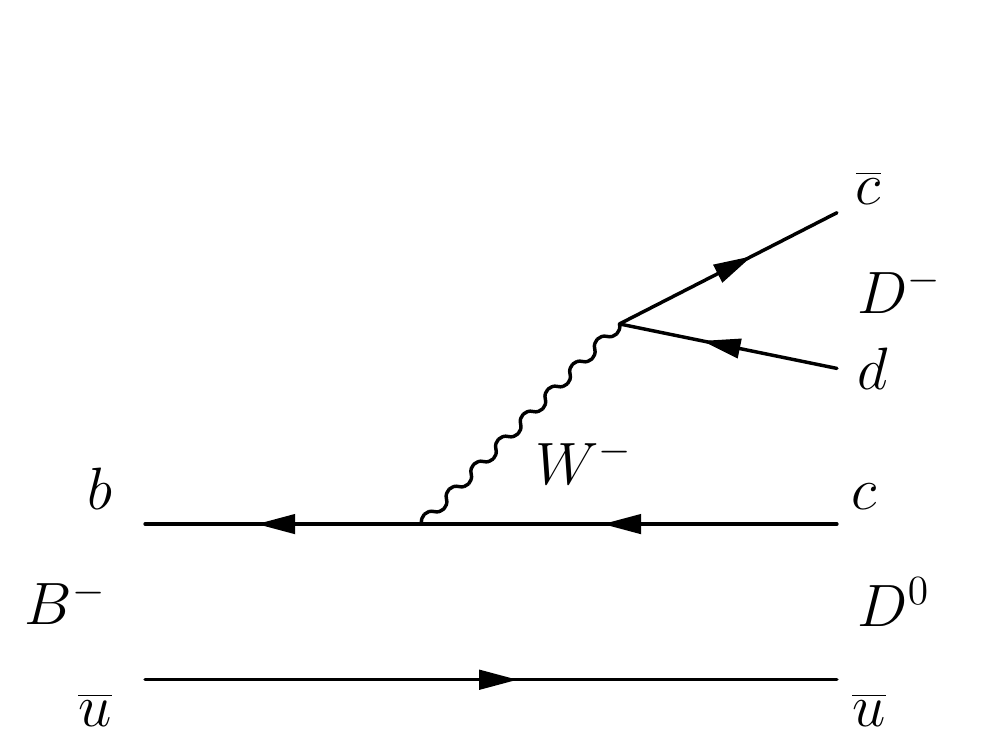}
\includegraphics[width=5cm]{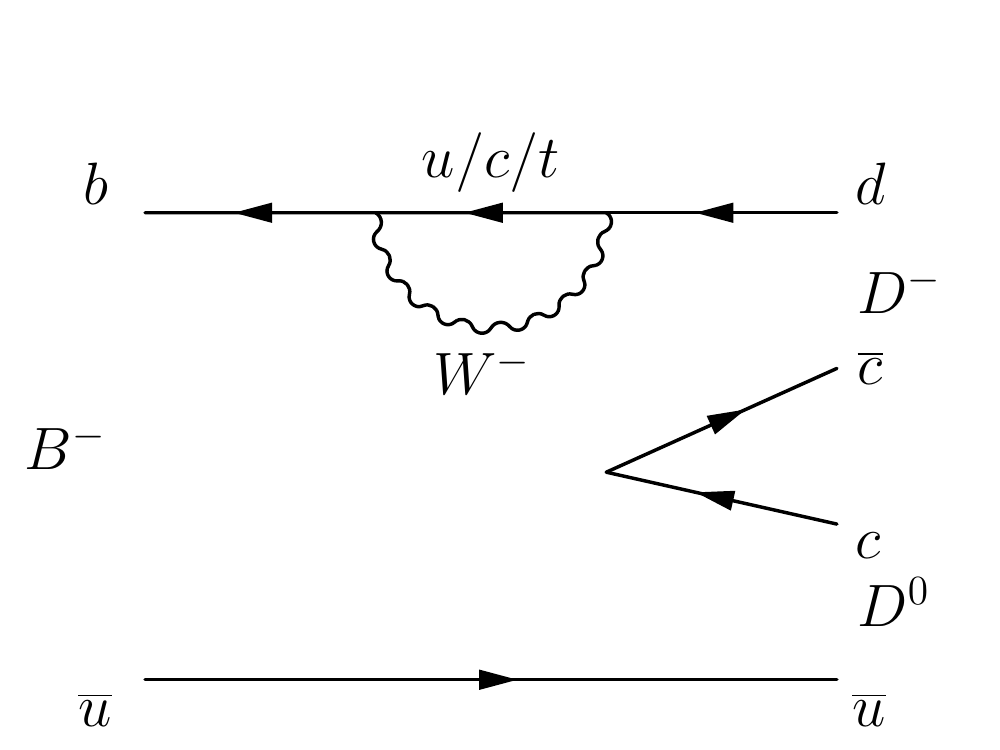}
\includegraphics[width=5cm]{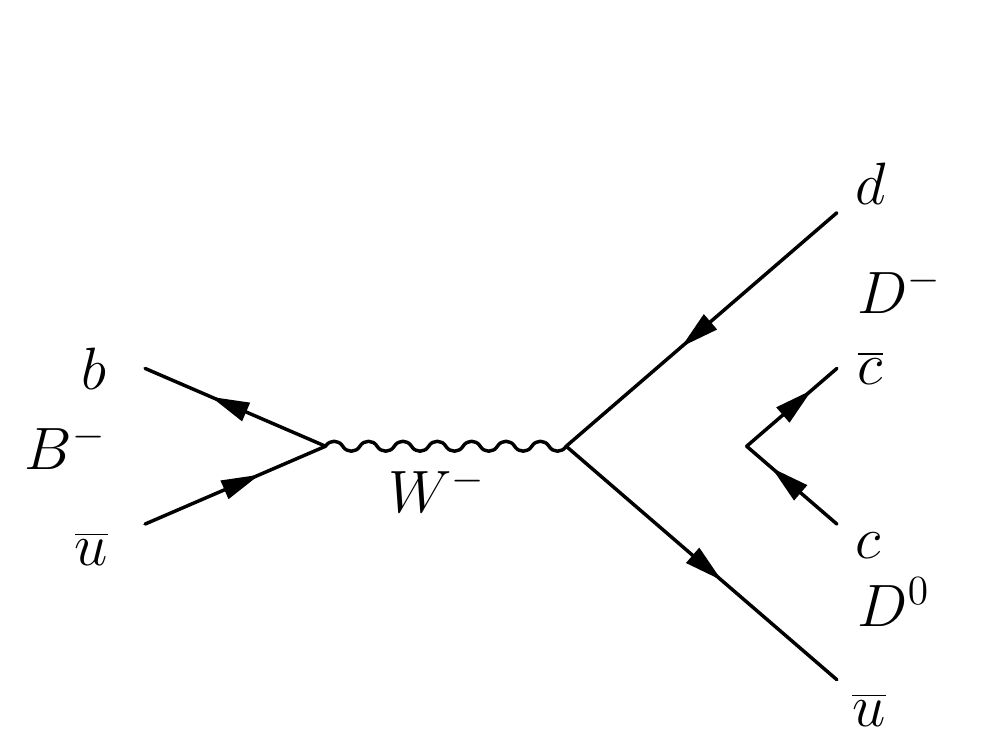}
\caption{Illustration of (left) tree, (centre) loop and (right) annihilation diagram contributions to the decay \decay{\Bm}{\Dm\Dz} in the SM.
Similar diagrams, with the \dquark quark replaced by an \squark quark, apply to the decay \decay{\Bm}{\Dsm\Dz}.
These diagrams also apply to decays with excited charm mesons.}
\label{fig:Diagrams_BpDpDzb}
\end{figure}

Interference of dominant tree-level amplitudes with sub-dominant loop-level and annihilation amplitudes (Fig.~\ref{fig:Diagrams_BpDpDzb}) produces nonzero direct \CP asymmetries in the decays of \B mesons to two charm mesons.
These are predicted to be small in the SM, up to 1\% for decays involving \mbox{\decay{\bquark}{\cquark\cquarkbar\squark}} transitions and up to 5\% for decays involving \mbox{\decay{\bquark}{\cquark\cquarkbar\dquark}} transitions~\cite{Kim:2008ex,Li:2009xf,Lu:2010gg,Xu:2016hpp}, as shown in Table~\ref{tab:ACPpred}.
These values may be enhanced by poorly understood QCD penguin diagrams~\cite{Jung:2014jfa} or contributions from models beyond the SM (BSM) such as supersymmetry~\cite{Kim:2008ex}, supergravity~\cite{Lu:2010gg} and a fourth generation of quarks~\cite{Xu:2016hpp}.
A combined analysis of \CP asymmetries and branching fractions would help to discriminate between potential BSM effects and enhanced QCD penguin contributions.

This paper describes a measurement of the \CP asymmetry of the seven \mbox{\decay{\Bm}{\DmorDsorDstar\DzorDstar}} decays shown in Table~\ref{tab:ACPpred}. 
In addition, measurements are presented of the branching fraction ratios
\begin{equation}
    R(\Dm\Dz/\Dsm\Dz) \equiv \frac{\BF(\decay{\Bm}{\Dm\Dz})}{\BF(\decay{\Bm}{\Dsm\Dz})} \frac{\BF(\decay{\Dm}{\Kp\pim\pim})}{\BF(\decay{\Dsm}{\Kp\Km\pim})}
\end{equation}
and
\begin{equation}
    R(\Dstarm\Dz/\Dm\Dz) \equiv \frac{\BF(\decay{\Bm}{\Dstarm\Dz})}{\BF(\decay{\Bm}{\Dm\Dz})}\frac{\BF(\decay{\Dstarm}{\Dzb\pim})\BF(\decay{\Dzb}{\Kp\pim})}{\BF(\decay{\Dm}{\Kp\pim\pim})}.
\end{equation}

\begin{table}[tb]
\centering
\caption{Predictions in the SM (first four columns) and world average measurements (final column)  of \CP asymmetries for \Bm decays to two charm mesons in percent.}
\label{tab:ACPpred}
\begin{tabular}{lr@{\:$\pm$\:}lccr@{\:$\pm$\:}lr@{\:$\pm$\:}l}
\hline
Decay & \multicolumn{2}{c}{Ref.~\cite{Kim:2008ex}} & Ref.~\cite{Li:2009xf} & Ref.~\cite{Lu:2010gg} & \multicolumn{2}{c}{Ref.~\cite{Xu:2016hpp}} & \multicolumn{2}{c}{Measured~\cite{PDG2022}}\\
\hline \vspace{-0.45cm} \\
\decay{\Bm}{\Dsm\Dz} & $-0.28$ & $0.06$ & - & $-0.26^{+0.05}_{-0.04}$ & $-0.14$ & $0.25$ & $-0.4$ & $0.7$\\
\decay{\Bm}{\Dssm\Dz} & $-0.065$ & $0.005$ & - & $-0.07^{+0.03}_{-0.02}$ & $-0.01$ & $0.10$ & \multicolumn{2}{c}{-} \\
\decay{\Bm}{\Dsm\Dstarz} & $0.045$ & $0.015$ & - & $\phantom{-}0.03^{+0.02}_{-0.02}$ & $0.08$ & $0.03$ & \multicolumn{2}{c}{-}\\
\decay{\Bm}{\Dm\Dz} & $4.95$ & $1.08$ & $\phantom{-}0.6^{+0.6}_{-0.1}$ & $\phantom{-}4.4^{+1.1}_{-0.4}$ & \multicolumn{2}{c}{-} & 1.6 & 2.5\\
\decay{\Bm}{\Dm\Dstarz} & $-0.80$ & $0.35$ & $-0.5^{+0.1}_{-0.4}$ & $-0.6^{+0.4}_{-0.2}$ & \multicolumn{2}{c}{-} & $13$ & $18$\\
\decay{\Bm}{\Dstarm\Dz} & $1.19$ & $0.16$ & $\phantom{-}0.1^{+0.6}_{-0.1}$ & $\phantom{-}1.2^{+0.4}_{-0.3}$ & \multicolumn{2}{c}{-} & $-6$ & $13$\\
\decay{\Bm}{\Dstarm\Dstarz} & $1.19$ & $0.16$ & $\phantom{-}0.2^{+0.0}_{-0.1}$ & $\phantom{-}1.2^{+0.4}_{-0.3}$ & \multicolumn{2}{c}{-} & $-15$ & $11$\\
\hline
\end{tabular}
\end{table}

The most precise measurements of the \CP asymmetries in \mbox{\decay{\Bm}{\DmorDs\Dz}} decays use data collected by the \lhcb experiment at centre-of-mass energies of
$\sqrt{s}=7$\tev and 8\tev corresponding to an integrated luminosity of $3\invfb$ of proton-proton $(pp)$ collisions~\cite{LHCb-PAPER-2018-007}.
Measurements of the \CP asymmetries for \mbox{\decay{\Bm}{\DmorDstar\DzorDstar}} decays are provided by the \belle and \babar experiments~\cite{BaBar:2006uih,Adachi:2008cj}.
No previous measurements of \mbox{$\ACP(\decay{\Bm}{\Dsm\Dstarz})$} and \mbox{$\ACP(\decay{\Bm}{\Dssm\Dz})$} exist.
The branching fractions \mbox{$\BF(\decay{\Bm}{\Dsm\Dz})=(9.0\pm0.9)\times10^{-3}$}, \mbox{$\BF(\decay{\Bm}{\Dm\Dz})=(3.8\pm0.4)\times10^{-4}$} and \mbox{$\BF(\decay{\Bm}{\Dstarm\Dz})=(3.9\pm0.5)\times10^{-4}$} have been measured by \belle~\cite{Adachi:2008cj,Belle:2005wxx} and \babar~\cite{BaBar:2006jvx,Aubert:2006ia}; \lhcb has measured the branching fraction of \decay{\Bm}{\Dsm\Dz} decays relative to \decay{\Bz}{\Dsm\Dp} decays~\cite{LHCb-PAPER-2012-050}.
The world average values are \mbox{$R(\Dm\Dz/\Dsm\Dz)=(7.4\pm1.1)\times10^{-2}$} and \mbox{$R(\Dstarm\Dz/\Dm\Dz)=0.29\pm0.05$}~\cite{PDG2022}.

The measurements presented in this paper use $pp$ collision data collected by the LHCb experiment, corresponding 
to an integrated luminosity of $9\invfb$, 
of which $1\invfb$ was recorded at centre-of-mass energy 
$\sqrt{s}=7$\tev, $2\invfb$ at $\sqrt{s}=8$\tev and $6\invfb$ at $\sqrt{s}=13$\tev.
The data taken at 7 and $8\tev$ are referred to as Run~1, and the data taken at $13\tev$ as Run~2.

Decays are reconstructed in the \mbox{$\Dsm\Dz$}, \mbox{$\Dm\Dz$} or \mbox{$\Dstarm\Dz$} final states.
Decays involving one \Dssm or \Dstarz meson are partially reconstructed with a photon or neutral pion that is not reconstructed.
These decays manifest themselves as broad structures in the invariant mass distributions of the charm meson candidate pairs, below the \Bm mass.
Charm mesons are reconstructed in the 
\mbox{\decay{\Dz}{\Km\pip}},
\mbox{\decay{\Dz}{\Km\pip\pim\pip}},
\mbox{\decay{\Dm}{\Kp\pim\pim}}, 
\mbox{\decay{\Dsm}{\Km\Kp\pim}} and
\mbox{\decay{\Dstarm}{\Dzb\pim}}
decay modes.
In the \mbox{\decay{\Bm}{\Dstarm\Dz}} final state, 
at least one of the neutral charm mesons
is required to decay as \mbox{\decay{\Dz}{\Km\pip}}.
When measuring \mbox{$R(\Dstarm\Dz/\Dm\Dz)$}, only decays of the \Dzb from the \Dstarm to $\Kp\pim$ are considered to minimise uncertainties from differences between the $\Dstarm\Dz$ and $\Dm\Dz$ decay modes.

The measurements of \mbox{$\ACP(\decay{\Bm}{\DmorDsorDstar\DzorDstar})$} are determined from the raw asymmetries
\begin{equation}
\Araw\equiv\frac{N(\decay{\Bm}{\DmorDsorDstar\DzorDstar})-N(\decay{\Bp}{\DporDsorDstar\DzborDstar})}{N(\decay{\Bm}{\DmorDsorDstar\DzorDstar})+N(\decay{\Bp}{\DporDsorDstar\DzborDstar})},
\label{eq:rawACP}
\end{equation}
where $N$ indicates the observed, uncorrected yield in the respective decay channel.
The raw asymmetries differ from the \CP asymmetries due to charge asymmetries in production of the \Bm meson and detection of the final states.
Since the asymmetries are small, higher-order terms corresponding to products of the asymmetries can be neglected, and the following relation holds
\begin{equation}
\ACP=\Araw-\AP-\AD,
\label{eq:ACP}
\end{equation}
where \AP is the asymmetry in the production cross-sections, $\sigma$, of \Bpm mesons,
\begin{equation}
\AP\equiv\frac{\sigma(\Bm)-\sigma(\Bp)}{\sigma(\Bm)+\sigma(\Bp)},
\label{eq:A_P}
\end{equation}
and \AD is the asymmetry of the detection efficiencies, $\varepsilon$,
\begin{equation}
\AD\equiv\frac{\varepsilon(\decay{\Bm}{\DmorDsorDstar\DzorDstar})-\varepsilon(\decay{\Bp}{\DporDsorDstar\DzborDstar})}{\varepsilon(\decay{\Bm}{\DmorDsorDstar\DzorDstar})+\varepsilon(\decay{\Bp}{\DporDsorDstar\DzborDstar})}.
\label{eq:A_D}
\end{equation}
Production and detection asymmetries are evaluated with measurements from calibration data samples that are corrected to match the kinematics of the signal decays.

Ratios of branching fractions are measured for the fully reconstructed decays where high precision is achievable.
These ratios are derived from the ratio of the efficiency-corrected yields,
\begin{equation}
    R(\Dm\Dz/\Dsm\Dz) = \frac{N(\decay{\Bm}{\Dm\Dz})}{N(\decay{\Bm}{\Dsm\Dz})} \frac{\varepsilon(\decay{\Bm}{\Dsm\Dz})}{\varepsilon(\decay{\Bm}{\Dm\Dz})}.
\end{equation}
The detection efficiency $\varepsilon$ includes the acceptance, reconstruction and selection efficiencies, and is determined using simulated samples of signal decays.
The quantity \mbox{$R(\Dstarm\Dz/\Dm\Dz)$} is measured analogously.

\section{Detector and simulation}
\label{sec:Detector}

The \lhcb detector~\cite{LHCb-DP-2008-001,LHCb-DP-2014-002} is a single-arm forward
spectrometer covering the \mbox{pseudorapidity} range $2<\eta <5$,
designed for the study of particles containing \bquark or \cquark
quarks. The detector includes a high-precision tracking system
consisting of a silicon-strip vertex detector surrounding the $pp$
interaction region~\cite{LHCb-DP-2014-001}, a large-area silicon-strip detector located
upstream of a dipole magnet with a bending power of about
$4{\mathrm{\,Tm}}$ and three stations of silicon-strip detectors and straw
drift tubes~\cite{LHCb-DP-2013-003,LHCb-DP-2017-001}
placed downstream of the magnet.

The tracking system provides a measurement of the momentum, \ptot, of charged particles with
a relative uncertainty that varies from 0.5\% at low momentum to 1.0\% at $200\gevc$.
The minimum distance of a track to a primary $pp$ collision vertex (PV), the impact parameter, 
is measured with a resolution of $(15+29/\pt)\mum$,
where \pt is the component of the momentum transverse to the beam, in\,\gevc.
Different types of charged hadrons are distinguished using information
from two ring-imaging Cherenkov detectors~\cite{LHCb-DP-2012-003}. 
Photons, electrons and hadrons are identified by a calorimeter system consisting of
scintillating-pad and preshower detectors, an electromagnetic
calorimeter (ECAL) and a hadronic calorimeter (HCAL). Muons are identified by a
system composed of alternating layers of iron and multiwire
proportional chambers~\cite{LHCb-DP-2012-002}.
The online event selection is performed by a trigger~\cite{LHCb-DP-2012-004}, 
which consists of a hardware stage, based on information from the calorimeter and muon
systems, followed by a software stage, which applies a full event
reconstruction.

At the hardware trigger stage, events are required to have a muon with high \pt or a
  hadron, photon or electron with high transverse energy in the calorimeters. For hadrons,
  the typical transverse energy threshold is $3.5\gev$.
  Signal candidates may be accepted if the candidate itself causes a positive decision for the hadron trigger,
hereafter called trigger on signal (TOS),
or due to the other particles produced in the $pp$ collision,
hereafter called trigger independent of signal (TIS).
  The software trigger requires a two-, three- or four-track
  secondary vertex with a significant displacement from any PV. 
  At least one track should have $\pt>1.7\gevc$ and \chisqip with respect to any
  PV greater than 16, where \chisqip is defined as the
  difference in the vertex-fit \chisq of a given PV reconstructed with and
  without the considered particle.
  A multivariate algorithm~\cite{BBDT,LHCb-PROC-2015-018} is used for
  the identification of secondary vertices consistent with the decay
  of a \bquark hadron.

Simulated events are used for the evaluation of signal efficiencies and in the training of multivariate classifiers.
Additionally, simulation of the invariant-mass and kinematic distributions of signal candidates are important for measuring the raw asymmetry and determining the production and detection asymmetries, respectively.
In the simulation, $pp$ collisions are generated using
\pythia~\cite{Sjostrand:2007gs,*Sjostrand:2006za} 
 with a specific \lhcb
configuration~\cite{LHCb-PROC-2010-056}.
Decays of hadronic particles
are described by \evtgen~\cite{Lange:2001uf}, in which final-state
radiation is generated using \photos~\cite{davidson2015photos}. The
interaction of the generated particles with the detector, and its response,
are implemented using the \geant
toolkit~\cite{Allison:2006ve, *Agostinelli:2002hh} as described in
Ref.~\cite{LHCb-PROC-2011-006}.
The simulated \Bm production cross-section is corrected to match the observed spectrum
  of \mbox{\decay{\Bm}{\Dsm\Dz}} decays in data where the background is subtracted using the \mbox{$m(\Dsm\Dz)$} sidebands.
  The correction is applied as a function of the transverse momentum and rapidity of the \Bm meson and the number of tracks in the event, using a gradient boosted reweighter~\cite{Rogozhnikov:2016bdp} technique.
The weights are determined separately for Run~1 and Run~2.
In the cases where the \mbox{\decay{\Dz}{\Km\pip\pim\pip}} decay is simulated according to a uniform distribution in the phase space, a gradient boosted reweighter is used to correct the simulation as a function of the two- and three-body invariant mass combinations of the \Dz decay products to match background-subtracted data.
In addition, corrections using control samples are applied to the simulated events
to improve the agreement with data regarding particle identification (PID) variables,
the momentum scale and the momentum resolution.
The visible momentum distribution of the \Bm meson is softer in decays with an unreconstructed particle.
For such decays, the kinematic distributions are estimated by weighting the simulated distributions of fully reconstructed decays by a  function of the form $p^{-\alpha}$ where the power is established using the observed function in data.

\section{Candidate selection}
\label{sec:selection}

Charm meson candidates are reconstructed by combining 2, 3 or 4 good-quality final-state tracks that have a significant impact parameter with all reconstructed PVs.
The tracks are required to form a high-quality vertex and the scalar sum of the \pt of the tracks must exceed 1.8\,\gevc.
To reduce background from misidentified particles, the pion and kaon candidates must also satisfy loose criteria on \dllkpi, the ratio of the likelihood between the kaon and pion PID hypotheses.
Since the detection asymmetry of kaons is large and the interaction cross-section of charged kaons with matter varies rapidly at low momenta~\cite{PDG2022}, all kaons are required to have momenta greater than 3.0\,\gevc.
The reconstructed mass of \Dz, \Dsm and \Dm candidates is required to be within $25\mevcc$ of their known values~\cite{PDG2022}.
For channels with a fully reconstructed \Dstarm meson that decays to \Dzb\pim, the mass difference $\Delta m$ between the \Dstarm and the \Dzb candidates is required to be within $10\mevcc$ of the known value~\cite{PDG2022}.
If more than one charm-meson candidate is formed from the same combination of tracks, only the most compatible with the charm-meson hypothesis according to the \dllkpi of the tracks and, for \DmorDs candidates, the consistency of the mass of two opposite-charged tracks with that of the $\phi(1020)$ meson, is selected.

In events with at least one \DmorDsorDstar candidate and at least one \Dz candidate,
the charm mesons are combined to form a \Bm candidate if the combination has a \pt greater than $4.0\gevc$, forms a good-quality vertex and points back to a PV. 
The reconstructed decay time of the charm meson candidates with respect to the \Bm vertex divided by its uncertainty, $t/\sigma_t$,
 is required to exceed $-3$ for \Dsm and \Dz mesons.
This requirement is increased to $+3$ for the longer-lived \Dm meson to eliminate background from \mbox{\decay{\Bm}{\Dz\pim\pip\pim}} decays where the positively charged pion is misidentified as a kaon.
Candidate \Bm decays that are compatible with the combination of
a \decay{\B^0}{\DmorDstar\pip(\pim\pip)} or \decay{\Bs}{\Dsm\pip(\pim\pip)} decay with one or three charged tracks, according to the invariant mass of a subset of the tracks forming the candidate, are rejected.
To eliminate duplicate tracks, the opening angle between any pair of final-state particles is required to be at least 0.5\,\mrad.
The invariant-mass resolution of \Bm decays is significantly improved by applying a kinematic fit~\cite{Hulsbergen:2005pu}.
This fit constrains the invariant masses of the \Dz and the \DmorDs candidates to their known values~\cite{PDG2022},
all trajectories of the decay products from the \DmorDs, \Dz, \Dstarm and \Bm decays to originate from their corresponding decay vertex,
and the \Bm candidate to originate from the PV with which it has the smallest \chisqip.

To reduce the combinatorial background while maintaining high signal efficiency,
a multivariate selection based on a boosted decision tree (BDT)~\cite{Breiman,Roe} is employed.
The BDT classifiers exploit kinematic and PID properties of selected candidates, namely:
the fit quality of the \Bm, \DmorDs and \Dz candidate decay vertices;
the value of \chisqip of the \Bm candidate;
the values of $t/\sigma_t$ of the \Bm, \DmorDs and \Dz candidates;
the unconstrained reconstructed masses of the \DmorDs and \Dz candidates;
the $\Delta m$ for the \Dstarm candidate;
and the reconstructed masses of the pairs of opposite-charge tracks from the \DmorDs candidate.
In addition, for each \DmorDs and \Dz candidate,
the smallest value of \pt and the smallest value of \chisqip among the decay products
and the smallest (largest) value of \dllkpi among all kaon (pion) candidates, are included as input variables for the BDT classifiers.

The BDT classifier for the branching fraction measurement is trained with a reduced set of variables that are accurately modelled in simulation or whose data-simulation differences can be corrected using the Cabbibo-favoured \mbox{\decay{\Bm}{\Dsm\Dz}} decay. 
The variables removed are:
the vertex fit quality and $t/\sigma_t$ for the \DmorDs candidates and \Dzb candidates from a \Dstarm; the reconstructed masses of the pairs of opposite-charge tracks from the \DmorDs candidate; the $\Delta m$ of the \Dstarm candidate; and the values of \dllkpi.

The BDT classifiers are trained separately for the \Dsm\Dz, \Dm\Dz and \Dstarm\Dz final states, for the \DKpi and \DKpipipi decay channels and for the Run~1 and Run~2 data samples.
The BDT training uses corrected simulated signal samples and data in the upper mass sideband of the \Bm meson ($5350<m(\DmorDsorDstar\Dz)<6200\mevcc$) as background.
To increase the size of the background sample, the charm meson invariant-mass intervals are increased from $\pm25\mevcc$ to $\pm75\mevcc$ and so-called wrong-sign \decay{\Bm}{\DmorDsorDstar\Dzb} candidates are also included.
Five-fold cross-training~\cite{geisser} is used to avoid biases in the calculation of the output of the BDT classifiers.
No dependence of the BDT classifiers on candidate charge or magnet polarity is observed.

The BDT classifiers combines all input variables into a single discriminant.
The optimal requirement on this value is determined by maximizing $N_S/\sqrt{N_S+N_B}$,
where $N_S$ is the expected signal yield, determined from the initial signal yield in data multiplied by the efficiency of the BDT requirement from simulation,
and $N_B$ is the background yield in a $\pm20\mevcc$ interval around the \Bm mass.
This selection has an efficiency of 87\% to 98\% and a background rejection of 81\% to 98\%.

Up to 3\% of events in data passing all selection and within $\pm40\mevcc$ of the \Bm mass contain more than one \Bp or \Bm candidate in a given decay channel.
For the measurement of \ACP, all candidates are retained to avoid biasing the detection asymmetries, while for measurements of the branching fractions, only one randomly selected candidate is retained per event.

\section{Measurement of the raw asymmetries and yields}
\label{sec:rawasym}

Raw asymmetries and yields are determined with an extended maximum-likelihood fit to the invariant-mass distribution of \decay{\Bm}{\DmorDsorDstar\Dz} candidates in the data.
The fit is performed in the range \mbox{$4870\leq m(\DmorDsorDstar\Dz)\leq 5400\mevcc$} in bins of width $1\mevcc$.
The fit is performed simultaneously on the \Bm and \Bp candidates and separate fits are performed for Run~1 and Run~2 data and for each \Dz meson decay mode.
The model includes: components for the signal decays;
decays of a \Bm, \Bz or \Bs meson to $\Dstar\DorDstar$ where one or both excited charm mesons emit a photon or pion that is not reconstructed;
partially reconstructed decays of a \Bm or \Bz meson to a $P$-wave charm meson excitation (referred to as \Dstst) plus another charm meson;
\mbox{\decay{\Bm}{\Dz\Km\Kp\pim}} decays;
cross-feed of Cabbibo-favoured \mbox{\decay{\Bm}{\Dsm\Dz}} decays to Cabbibo-suppressed \mbox{\decay{\Bm}{\Dm\Dz}} decays;
and the combinatorial background.

The invariant-mass distribution of \mbox{\decay{\Bm}{\DmorDsorDstar}{\Dz}} decays is described by the sum of a Gaussian function and a double sided Crystal Ball (DSCB) function, which is a Crystal Ball~\cite{Skwarnicki:1986xj} function extended to have power-law tails on both the low-mass and the high-mass sides.
The DSCB and Gaussian functions share a common peak position.
The tail parameters of the DSCB function and the ratio of the integrals of the DSCB and Gaussian components are fixed from simulation.
The ratio of the widths of the DSCB and Gaussian functions is constrained from simulation.
The peak position and the overall width are free parameters in the fit to data.

The invariant-mass distribution of \Bm, \Bz or \Bs meson decays to two charm mesons, where one of the excited charm mesons emits a pion or photon that is not reconstructed, is parameterised by a parabola convolved with a resolution function.
The parameters of the parabola depend analytically on the spin of the charm mesons and the missing particle, the polarisation of the vector states in pseudoscalar to vector vector decays and the masses of the particles in the decay chain~\cite{LHCb-PAPER-2017-021,LHCb-PAPER-2016-006}.
The parabola endpoints, which depend on the particle masses, are allowed to vary from their expected values by a shift that is shared with the shift of the fully reconstructed signal peak position from the \Bm meson mass.
The parabola is multiplied by a linear function~\cite{LHCb-PAPER-2017-021} with gradient shared between all decays to the same final state, to account for the dependence of reconstruction and selection efficiencies on the invariant mass.
The product is convolved with a resolution function that is the sum of four Gaussian functions, of which two act as the core of the resolution function and two act as the small high- and low-mass tails.
The integral of the narrower core Gaussian component and the width ratio between the two core Gaussian components are fixed to match the integral of the Gaussian component and the width ratio between the Gaussian and DSCB components in the fully reconstructed signal model.
The tail parameters are fixed using a fit of the resolution model to fully reconstructed signal simulation.

Decays of \Bm, \Bz or \Bs mesons to two excited charm mesons (denoted as \mbox{$\decay{\B}{\Dstar\Dstar}$}), where both excited charm mesons emit an unreconstructed pion or photon, exhibit a more complicated invariant mass distribution.
These distributions are obtained using simulation where the detector resolution is obtained from simulated fully reconstructed \mbox{\decay{\Bm}{\DmorDsorDstar}{\Dz}} decays.
Kernel fits~\cite{Cranmer:2001} to this simulation, multiplied by a linear function to include the dependence of efficiency upon invariant mass, are used to model these decays.
The linear gradient is shared between all such decays contributing to a final state.

Separate simulation samples are generated for each \Dstarm, \Dstarz and \Dssm decay and for both longitudinal and transverse polarisations in pseudoscalar decays to two vectors.
The relative yield corresponding to each \Dstar decay is fixed according to known branching fractions~\cite{PDG2022}.
The longitudinal polarisation fraction corresponding to the \mbox{\decay{\Bm}{\Dstarm\Dstarz}} signal decay varies freely in the \mbox{$\Dstarm\Dz$} final state to avoid a strong dependence of the results on theory predictions.
Elsewhere, polarisation fractions are fixed to measured values or theory predictions~\cite{PDG2022,LHCb-PAPER-2021-006,Belle:2012xkw,Kim:2008ex,Li:2009xf,Lu:2010gg,Xu:2016hpp}.
No measurements or predictions of \mbox{$f_L(\decay{\Bs}{\Dstarm\Dstarp})$} are available, so this parameter is assigned a value of 50\% in the baseline model.
For decays with two missing particles, the relative yields for the \Bm and \Bz decays are fixed using the measured branching fractions~\cite{PDG2022,BaBar:2006uih,Belle:2012xkw} and assuming equal production of the \Bm and \Bz mesons.
The ratios between the yields of the \Bs and \Bz meson decays to the same final state are fixed according to: measured branching fractions~\cite{PDG2022,LHCB-PAPER-2020-037,LHCB-PAPER-2022-023} or predictions where measurements are not available~\cite{Kim:2008ex,Li:2009xf,Lu:2010gg}; the ratio of the \Bs to \Bz fragmentation fractions ($f_s/f_d$) ~\cite{LHCb-PAPER-2020-046}; and the efficiency ratio $\varepsilon(\Bs)/\varepsilon(\Bz)$, which is estimated from the range of values in studies of \mbox{\decay{\Bs}{\Dstarpm\DmporDstar}} decays at \lhcb~\cite{LHCB-PAPER-2020-037,LHCB-PAPER-2022-023}.

Decays of a \B meson to a \Dstst meson plus a fully reconstructed charm meson (denoted as \mbox{$\decay{\B}{\Dstst\D}$}) where the decay chain of the \Dstst contains one or more pions or photons that are not reconstructed, enter the fit region at low invariant mass.
The model of these decays includes the $D_0^*(2300)$, $D_1(2420)$, $D_1(2430)$, $D_2^*(2460)$, $D_{s0}^*(2317)$ and $D_{s1}(2460)$ mesons and all decay chains of each \Dstst that have non-negligible branching fractions.
The invariant-mass model for each decay is obtained in the same way as for \mbox{\decay{\B}{\Dstar\Dstar}} with two missing particles and shares the efficiency gradient.
The relative yield of each decay is fixed using measured \Dstst and \Dstar branching fractions~\cite{PDG2022,BaBar:2006jvx}, assuming in the baseline model equal branching fractions for each \mbox{\decay{\B}{\Dstst\D}} decay and equal production of \Bm and \Bz mesons.

The yield of the Cabibbo-favoured \mbox{\decay{\Bm}{\Dz\Km\Kp\pim}} decay is strongly suppressed by the invariant mass and $t/\sigma_t$ requirements on the $\Dsm$ meson, but it still forms a non-negligible background.
This background is modelled by a DSCB function.
The peak position is shared with the fully reconstructed signal.
Other shape parameters are determined from a fit to simulated decays and corrected for the difference in the resolution between data and simulation using fully reconstructed signal. 
The yields are determined from the \Dsm sidebands and are approximately 3\% of the fully reconstructed signal yield.
Cross-feed of Cabibbo-favoured \mbox{\decay{\Bm}{\Dsm\Dz}} decays to the $\Dm\Dz$ final state is suppressed by PID requirements, but simulated samples predict that the yield of this cross-feed is 4\% of the yield of the \mbox{\decay{\Bm}{\Dm\Dz}} decay.
This misidentified background is modelled by the sum of two Gaussian functions with a shared mean, and the shape parameters are obtained from a fit to simulated samples.
The combinatorial background is described by an exponential function.

To improve precision on the raw asymmetries of signal decays, the raw asymmetries of background processes are constrained wherever constraints are available from external inputs or other channels in this analysis.
The value of \Araw for the \mbox{\decay{\BdorBs}{\Dstarm\Dstarp}} decay, involving a \CP-symmetric final state, is constrained to the detection asymmetry of the partially reconstructed decay products (Sect.~\ref{sec:corrections}). 
The time-integrated raw asymmetry of the \mbox{\decay{\Bz}{\Dm\Dstarp}} decay is approximated as
\begin{equation}
\Araw(\decay{\Bz}{\Dm\Dstarp}) = \ACP(\decay{\Bz}{\Dm\Dstarp}) + \AD(\decay{\Bz}{\Dm D^{*+}_{[\Dz]}}),
\end{equation}
where \mbox{$\ACP(\decay{\Bz}{\Dm\Dstarp})$} has been measured~\cite{PDG2022} and \mbox{$\AD(\decay{\Bz}{\Dm\Dstarp_{[\Dz]}})$} is the detection asymmetry when the \Dstarp is partially reconstructed as a \Dz, and neglecting the $\mathcal{O}(0.1\%)$ contribution from time-dependent flavour oscillations convolved with the production asymmetry.
The raw asymmetry of the \mbox{$\decay{\Bz}{\Dsm\Dstarp}$} decay is constrained to
\begin{equation}
    \Araw(\decay{\Bz}{\Dsm\Dstarp}) = \ACP(\decay{\Bz}{\Dsm\Dstarp})+\AD(\decay{\Bz}{\Dsm D^{*+}_{[\Dz\pip]}})-\AD(\pip) + d\AP(\Bz)
\end{equation}
where the first two terms on the right hand side have been measured in Ref.~\cite{LHCb-PAPER-2019-036}, and $\AP(\Bz)$~\cite{LHCb-PAPER-2019-036} is multiplied by a dilution factor that is assigned a value of 0.5 in the baseline model to account for time-dependent flavour oscillations.
Theoretical predictions of \mbox{$\ACP(\decay{\Bm}{\Dssm\Dstarz})$} and \mbox{$\ACP(\decay{\BdorBs}{\DsmorDstar\Dstarp})$}~\cite{Li:2009xf,Lu:2010gg,Kim:2008ex,Xu:2016hpp} are not used to avoid a strong dependence of the results on Standard Model assumptions.
The precision on \mbox{$\ACP(\decay{\Bm}{\Dstarm\DzorDstar})$} in the $\Dm\Dz$ channel is much worse than in the $\Dstarm\Dz$ channel due to the lower branching fraction of \decay{\Dstarm}{\Dm\piz} compared to \decay{\Dstarm}{\Dz\pip} and the broader invariant-mass distribution resulting from an additional unreconstructed particle.
Therefore \mbox{$\ACP(\decay{\Bm}{\Dstarm\DzorDstar})$} are constrained in the $\Dm\Dz$ channel to measurements in the $\Dstarm\Dz$ channel.
The raw asymmetry of the \mbox{\decay{\Bm}{\Dsm\Dz}} cross-feed to the $\Dm\Dz$ channel is equal to the raw asymmetry in the $\Dsm\Dz$ channel.
The \CP asymmetries of Cabbibo-favoured single-charm \Bm meson decays are expected to be small,
since they are dominated by a single decay amplitude.
The value of \mbox{$\ACP(\decay{\Bm}{\Dz\Km\Kp\pim})$} has not been measured but \mbox{$\ACP(\decay{\Bm}{\Dz\pim\pip\pim})=(-0.2\pm1.1)\%$}~\cite{PDG2022}, which is a decay with the same flavour structure, has been measured.
Therefore, a \CP asymmetry with double the uncertainty, \mbox{$\ACP(\decay{\Bm}{\Dz\Km\Kp\pim})=(0.0\pm2.2)\%$} is assigned in the constraint of the raw asymmetry of this background.

\begin{figure}[p]
\includegraphics[width=8.0cm]{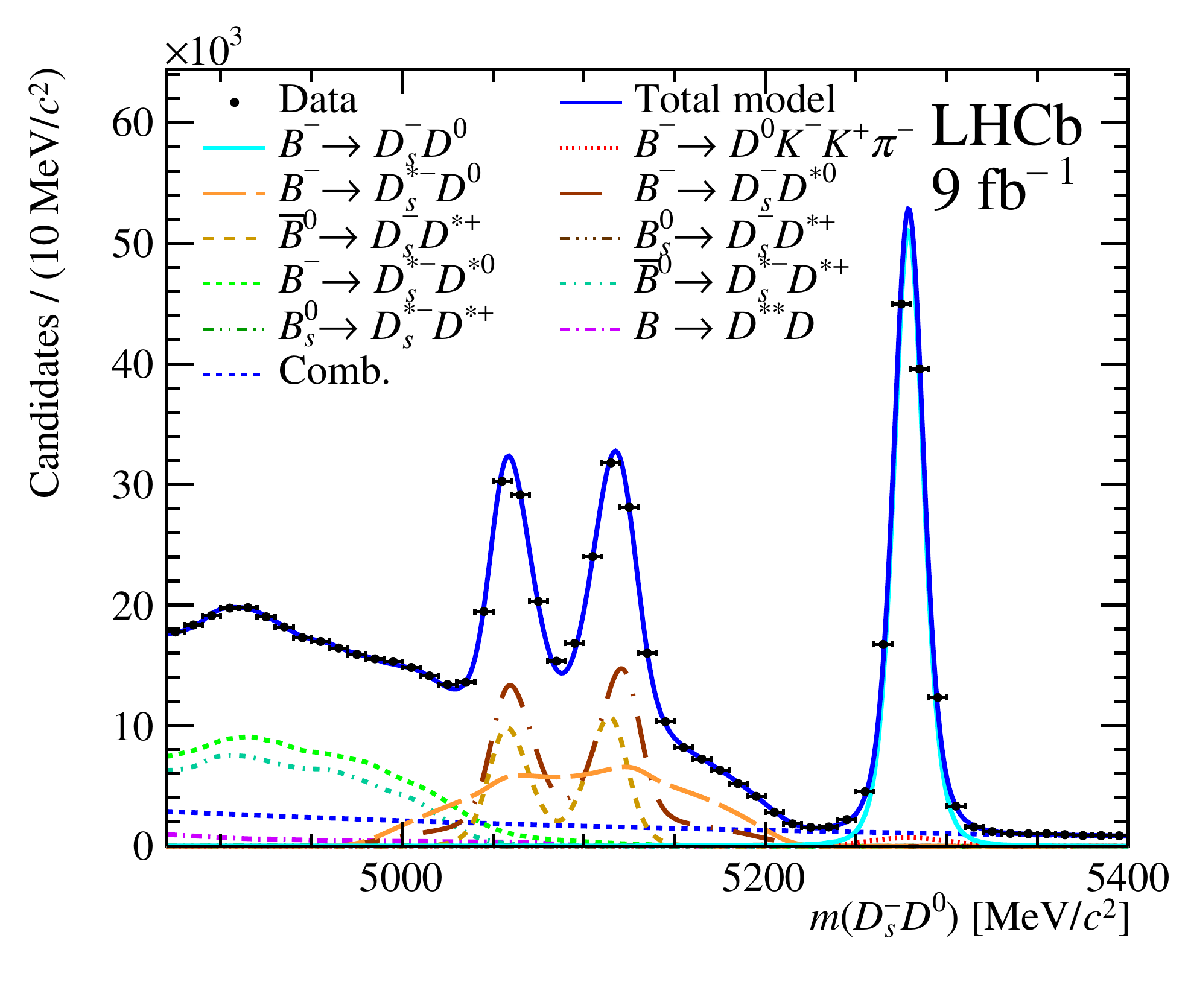}
\includegraphics[width=8.0cm]{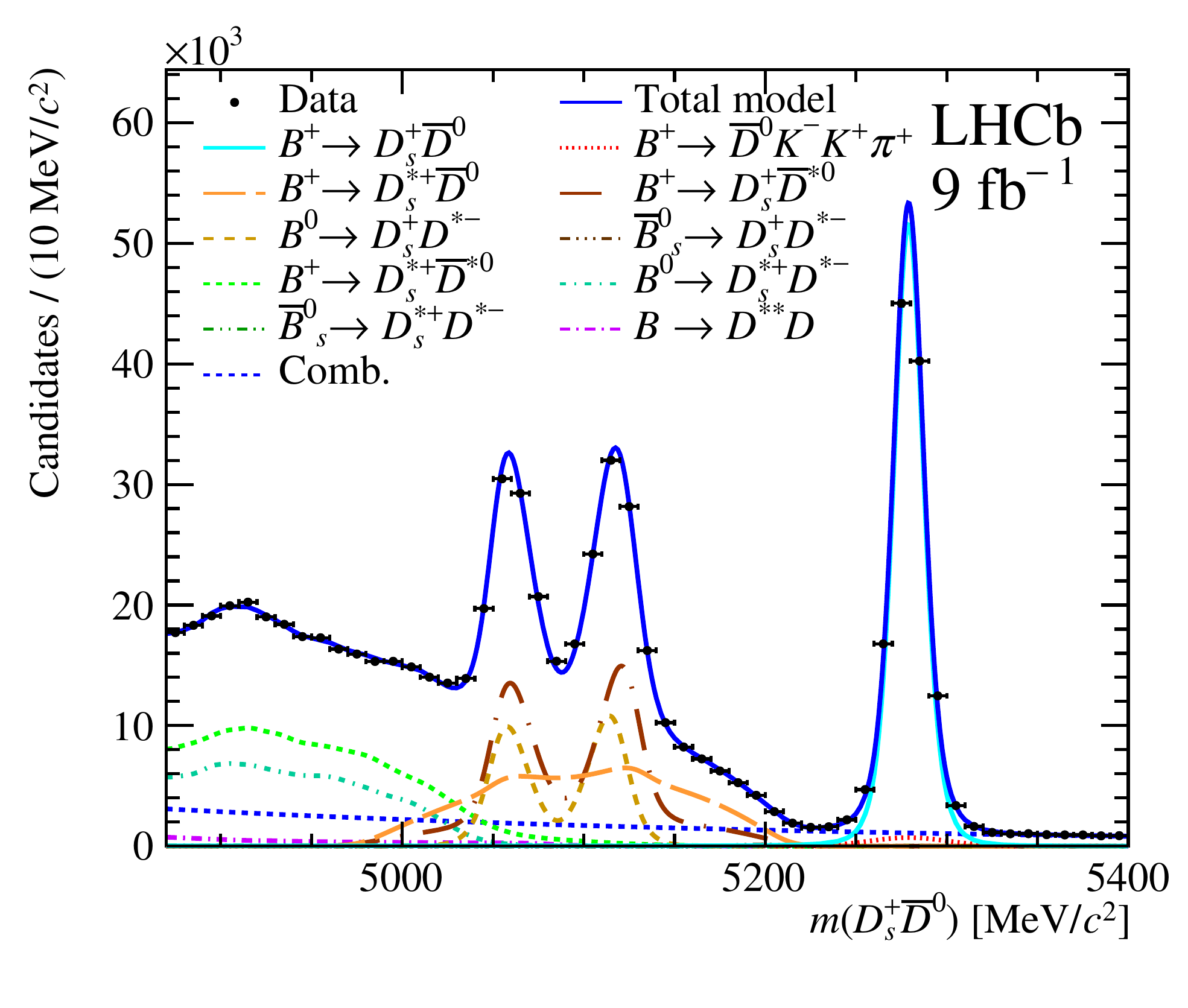}\\
\includegraphics[width=8.0cm]{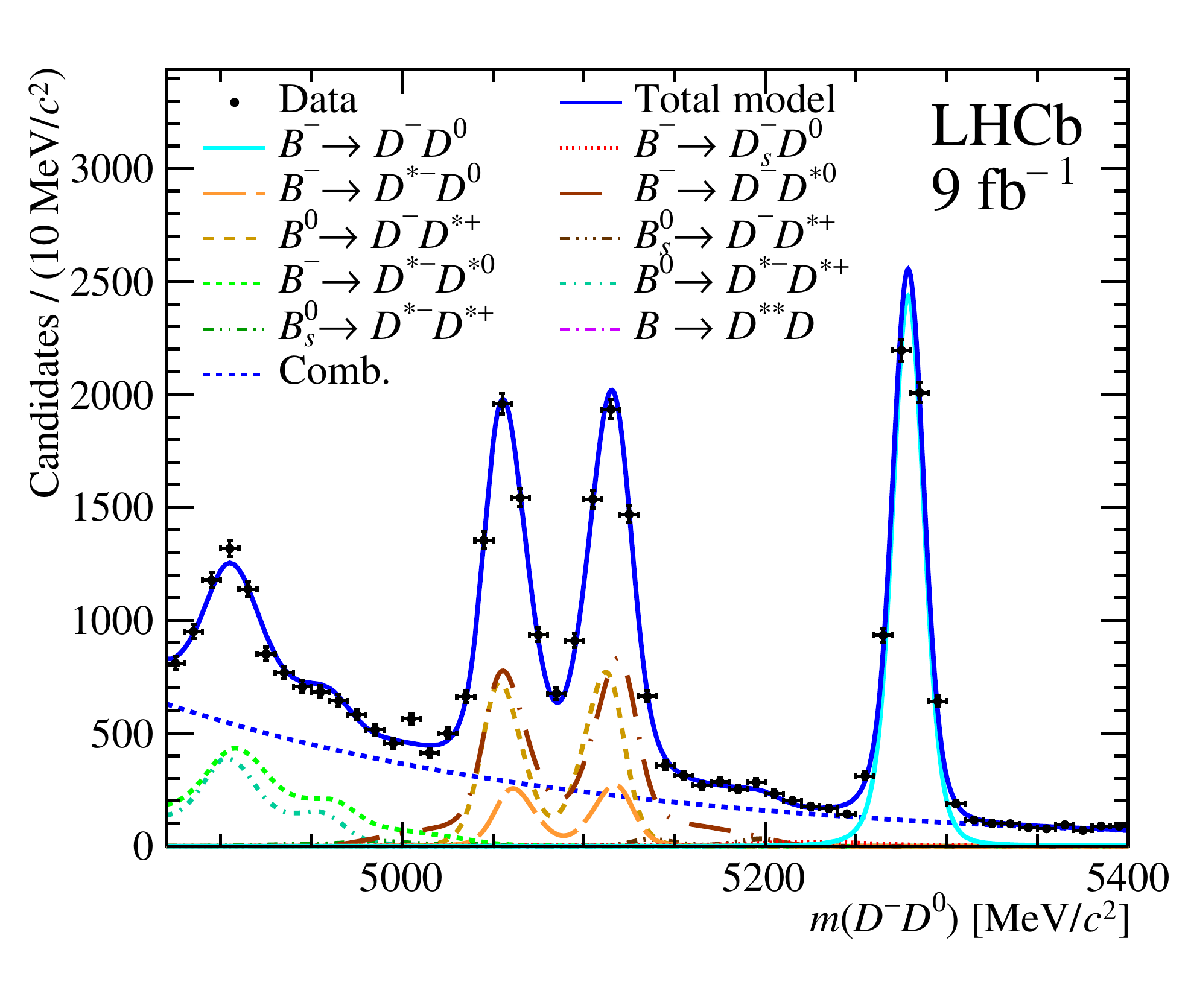}
\includegraphics[width=8.0cm]{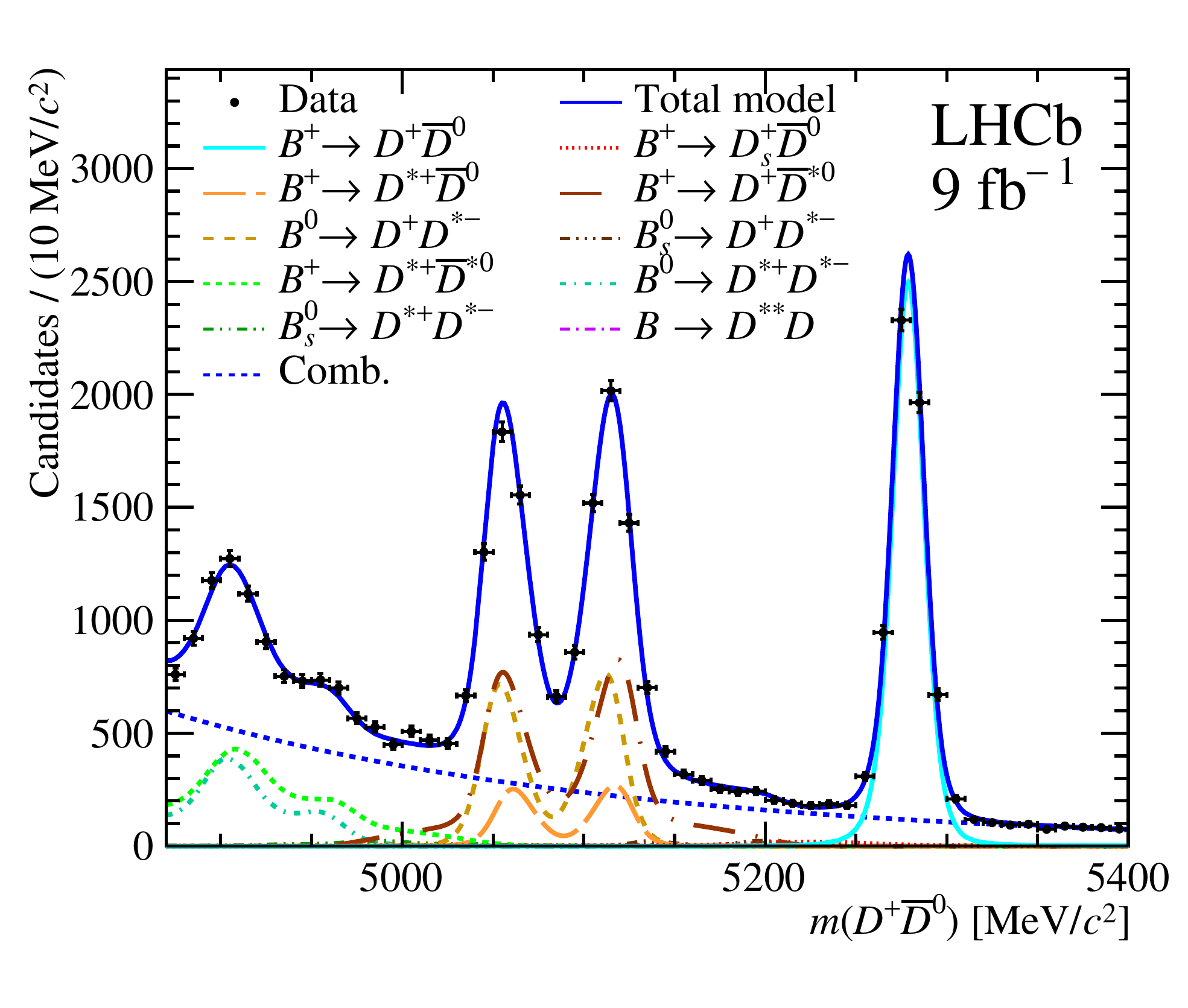}\\
\includegraphics[width=8.0cm]{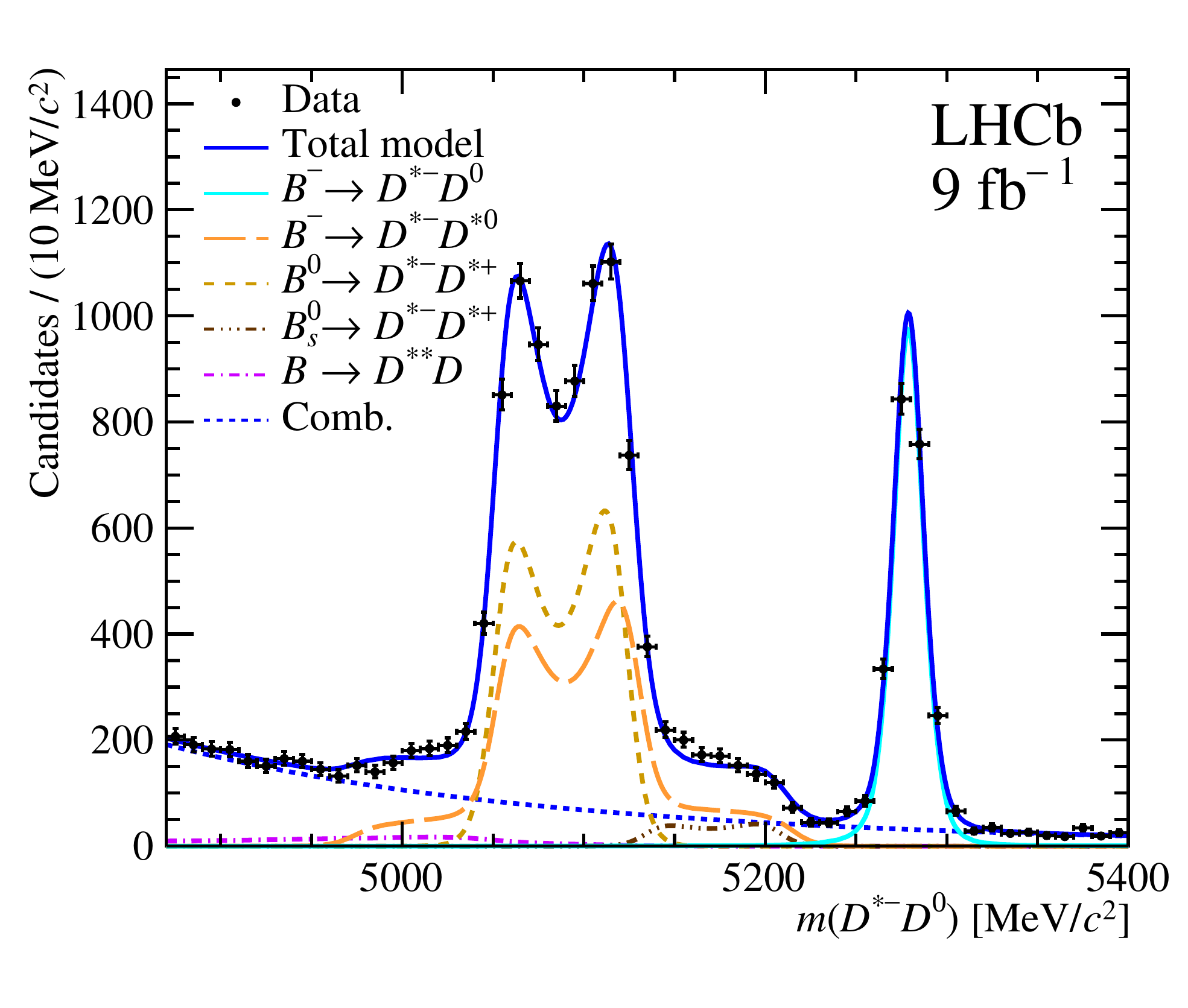}
\includegraphics[width=8.0cm]{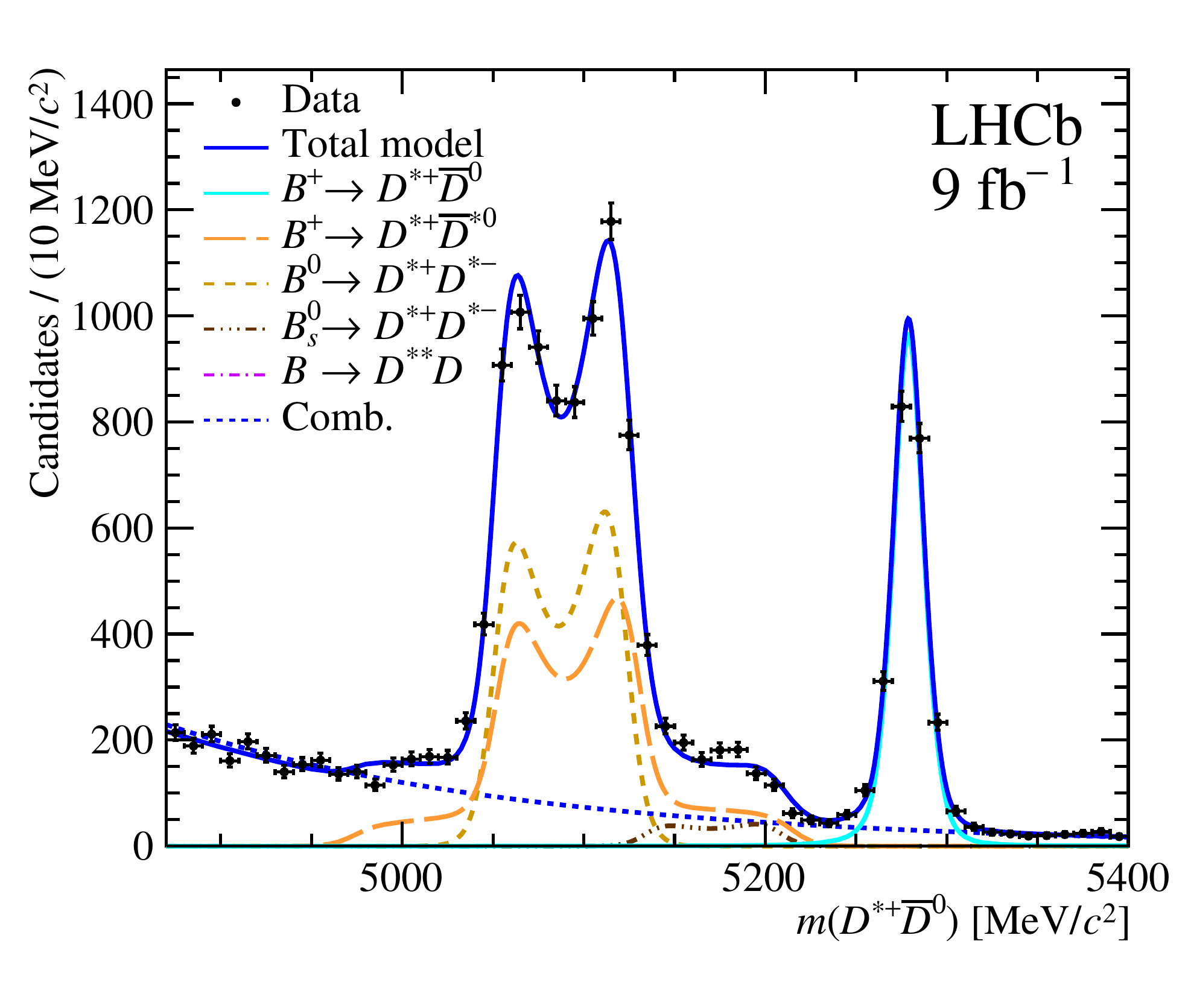}\\
\caption{Invariant-mass distribution of (top row) \mbox{\decay{\Bm}{\Dsm\Dz}} candidates, (middle row) \mbox{\decay{\Bm}{\Dm\Dz}} candidates and (bottom row) \mbox{\decay{\Bm}{\Dstarm\Dz}} candidates, separated by charge into (left) \Bm and (right) \Bp candidates.
The fitted model described in the text is also drawn.
The invariant-mass distributions of \mbox{\decay{\B}{\Dstst\D}} decays where the fitted yield is less than zero are not drawn.
}
\label{fig:Araw}
\end{figure}

The invariant mass distributions of the \Bpm candidates and the fitted models, summed over all centre-of-mass energies and over all \Dz decay modes, are shown in Fig.~\ref{fig:Araw}.
The charge-combined yields of the fully reconstructed decays in each channel, summing over all datasets, are $N(\Dsm\Dz)=230\,500 \pm 500$, $N(\Dm\Dz)=11\,490 \pm 120$ and $N(\Dstarm\Dz)=3\,100 \pm 70$.
The measured raw asymmetries, alongside the production and detection asymmetries that will be evaluated in the next section, are given in Table~\ref{tab:asymmetries}.

Ratios of branching fractions between decays with one missing particle and fully reconstructed decays measured from the above fits are in agreement with known values.
To assess bias on the raw asymmetries and the coverage of the statistical uncertainties, pseudoexperiments are generated according to Poisson statistics while accounting for the fraction of \mbox{\decay{\BdorBs}{\Dstarm\Dstarp}} decays that produce two candidates.
Biases in the raw asymmetries are at least five times smaller than the corresponding statistical uncertainties, no undercoverage is observed and overcoverage is below 20\%.

\boldmath
\section{Production and detection asymmetries}
\unboldmath
\label{sec:corrections}

The asymmetry of \Bm meson production has been measured by the \lhcb experiment as a function of the \Bm transverse momentum and rapidity and at centre-of-mass energies of 7, 8 and 13\tev using \mbox{\decay{\Bp}{\jpsi\Kp}} decays~\cite{LHCb-PAPER-2016-062,LHCb-PAPER-2021-049} with a measured value of \mbox{$\ACP(\decay{\Bp}{\jpsi\Kp})=(0.18\pm0.30)\%$}~\cite{PDG2022}.
The value of \AP for signal decays is determined by weighting these measurements by the kinematic distributions of selected signal candidates in simulation.
The asymmetry ranges from $-0.6\%$ to $-1.5\%$, depending on the centre-of-mass energy and the kinematic distribution of the signal decay.

The total detection asymmetry is the sum of all individual sources of detection asymmetry.
Asymmetries from the nuclear interaction of final-state particles with the detector material, track reconstruction and acceptance contribute to the detection asymmetry.
While the detection asymmetry is dominated by the \Km nuclear interaction asymmetry, the other sources produce small but non-negligible effects.
The difference in detection asymmetry between kaons and pions, 
\begin{equation}
    \AKpi \equiv \frac{\varepsilon(\Km\pip)-\varepsilon(\Kp\pim)}{\varepsilon(\Km\pip)+\varepsilon(\Kp\pim)}
\end{equation}
is determined by weighting the difference in raw asymmetry between prompt \mbox{\decay{\Dp}{\Km\pip\pip}} and \mbox{\decay{\Dp}{\Kzb\pip}} decays, corrected for the detection and \CP asymmetries of the \Kzb~\cite{LHCb-PAPER-2014-013,LHCb-PAPER-2021-016}, by the kinematic distribution of signal.
This asymmetry is between $-0.6\%$ and $-1.0\%$ per \Km.
The detection asymmetry for residual pions, 

\begin{equation}
    \Api \equiv \frac{\varepsilon(\pip)-\varepsilon(\pim)}{\varepsilon(\pip)+\varepsilon(\pim)}
\end{equation}
is determined by weighting an asymmetry determined by comparing the yields of fully to partially reconstructed \mbox{\decay{\Dstarp}{(\decay{\Dz}{\Km\pip\pim\pip})\pip}} decays~\cite{LHCb-PAPER-2016-013} by the momentum distribution of signal pions.
The asymmetry is up to $-0.4\%$ for the soft \pip from the \Dstarp decay, or up to $-0.2\%$ for other pions.

Particle identification information is used in the selection of candidates, including in the BDT classifiers.
Effective PID criteria are defined as the PID requirements that produce the same mean \dllkpi as the BDT requirement.
The charge asymmetry in the efficiency of the effective PID criteria, \APID, is determined as a function of track kinematics and event multiplicity from a sample of \mbox{\decay{\Dstarp}{(\decay{\Dz}{\Km\pip})\pip}} decays that are selected without PID requirements.
Since the BDT classifiers use PID information, the PID asymmetry is calculated using simulation without the BDT selection.
Charge asymmetries are also calculated for the PID criteria that are applied to the calibration samples used to evaluate \AKpi and \Api, which are often tighter than those applied to the signal candidates.
The largest values of \APID are induced by the comparatively tight PID requirements applied to the calibration samples used to evaluate \AKpi in Run~1 and are up to $\pm0.3\%$.

The hardware trigger requirements may also induce a charge asymmetry.
The TIS asymmetry, \ATIS, is independent of the signal decay but may depend on the $\pt(\Bm)$ and the magnet polarity and has been measured using \decay{\B}{\Dzb\mup\neum\PX} decays~\cite{LHCb-PAPER-2016-054}.
After weighting by the kinematics of the signal decay and averaging over magnet polarities, the asymmetry is evaluated to be \mbox{$(0.0\pm0.2)\%$}, where the uncertainty comes from the calibration sample size.
The efficiency of the hadron hardware trigger TOS requirement for pions and kaons of both charges has been measured as a function of transverse energy and position at the surface of the HCAL with a sample of \mbox{\decay{\Dstarp}{(\decay{\Dz}{\Km\pip})\pip}} decays~\cite{LHCb-PUB-2011-026}.
Corrections are applied for the overlapping energy deposits from other tracks in the decay and the underlying event, and relative miscalibration between HCAL cells.
The corresponding asymmetry, \ATOS, is determined in bins of \mbox{$(\pt(\Bm),\eta(\Bm))$} using an unbiased sample of simulated TIS signal candidates.
To determine the average value of \ATOS,
this binned measurement is weighted by the kinematics of 
simulated signal candidates passing the hadron hardware trigger TOS requirement.
The resulting asymmetry is larger for final states containing an odd number of kaons and is at most $-0.2\%$.
The corrections to the \CP asymmetries are obtained by multiplying each asymmetry by the fraction of selected candidates in background-subtracted data that pass the TIS or TOS requirements, which are both approximately 60\%.

The production and total detection asymmetries evaluated using this method are given for each decay in Table~\ref{tab:asymmetries}.

\begin{table}[tb]
\centering
\caption{Values of \Araw, \AP and \AD in percent, averaged over all \Dz decay modes and data-taking periods. The uncertainties on \Araw are statistical and systematic, respectively. The first uncertainty on \AP contains all sources of uncertainty except that on \mbox{$\ACP(\decay{\Bp}{\jpsi\Kp})$}, which is the second uncertainty.}\label{tab:asymmetries}
\begin{tabular}{lr@{\:$\pm$\:}r@{\:$\pm$\:}lr@{\:$\pm$\:}r@{\:$\pm$\:}lr@{\:$\pm$\:}l}
\hline
Decay & \multicolumn{3}{c}{\Araw} & \multicolumn{3}{c}{\AP}& \multicolumn{2}{c}{\AD} \\
\hline \vspace{-0.45cm} \\
$\Dsm\Dz$ & $-1.3$ & $0.2$ & $0.1$ & $-1.1$ & $0.3$ & 0.3 & $-0.7$ & $0.2$ \\
$\Dssm\Dz$ & $-2.4$ & $1.1$ & $0.9$ & $-1.1$ & $0.4$ & 0.3 & $-0.8$ & $0.2$ \\
$\Dsm\Dstarz$ & $-0.8$ & $0.8$ & $0.4$ & $-1.1$ & $0.4$ & 0.3 & $-0.8$ & $0.2$ \\
$\Dm\Dz$ & $1.5$ & $1.0$ & $0.2$ & $-1.1$ & $0.4$ & 0.3 & $0.1$ & $0.2$ \\
$\Dm\Dstarz$ & $-1.3$ & $2.0$ & $1.3$ & $-1.1$ & $0.4$ & 0.3 & $0.1$ & $0.2$ \\
$\Dstarm\Dz$ & $2.4$ & $1.6$ & $0.2$ & $-1.2$ & $0.4$ & 0.3 & $0.2$ & $0.3$ \\
$\Dstarm\Dstarz$ & $1.3$ & $2.1$ & $1.6$ & $-1.1$ & $0.5$ & 0.3 & $0.1$ & $0.2$ \\
\hline
\end{tabular}
\end{table}

\section{Efficiencies}
\label{sec:efficiency}

The detector acceptance and the efficiencies of final-state particle reconstruction and selection are evaluated using simulated samples.
These samples are corrected with per-candidate weights to better model the production of signal decays and the detector response.
In addition to the weights described in Sect.~\ref{sec:Detector} to correct the \Bm production distribution and the angular structure of the \decay{\Dz}{\Km\pip\pim\pip} decay, weights are applied to correct the efficiency of the hardware trigger, track reconstruction and PID criteria.
The efficiency of the hadron hardware trigger TOS requirement, $\varepsilon_{\rm TOS}$, is measured in bins of \mbox{$(\pt(\Bm),\eta(\Bm))$} as described in Sect.~\ref{sec:corrections}.
Per-bin weights that correct mismodelling of $\varepsilon_{\rm TOS}$ and the exclusively TIS (xTIS) efficiency, $\varepsilon_{\rm xTIS}=1-\varepsilon_{\rm TOS}$, for events passing the hardware trigger TIS requirement, are evaluated as the ratio between the measured and simulated efficiencies.
The appropriate weight is applied to each candidate, which modifies the hardware trigger efficiency by up to 3\%.
Weights to correct the tracking efficiency as a function of track momentum and pseudorapidity are determined using \mbox{\decay{\jpsi}{\mup\mun}} decays~\cite{LHCb-DP-2013-002}, and are applied to each final-state particle, resulting in a change of up to 6\%.
The efficiencies in data of PID requirements are derived as described in Sect.~\ref{sec:corrections}, and differ from the simulated efficiencies by up to 3\%.
Each efficiency correction is derived separately for each data-taking year and magnet polarity.

\section{Systematic uncertainties}
\label{sec:systematics}

An overview of the systematic uncertainties from all sources, averaged over \Dz decay modes, are listed in Tables~\ref{tab:systACPraw} and~\ref{tab:systACPcor} for the \CP asymmetry measurements and Table~\ref{tab:systBF} for the branching fraction measurements.
The following paragraphs describe the systematic effects and their evaluation in detail.

\begin{table}[tb]
    \caption{Systematic uncertainties on the raw asymmetries in percent, averaged over all \Dz decay modes.}
    \begin{center}
\begin{tabular}{lcccccc}
\hline
& Run~1 & Run~2 & Run~1 & Run~2 & Run~1 & Run~2 \\
\hline \vspace{-0.45cm} \\
& \multicolumn{2}{c}{$\decay{\Bm}{\Dsm\Dz}$} 
& \multicolumn{2}{c}{$\decay{\Bm}{\Dm\Dz}$} 
& \multicolumn{2}{c}{$\decay{\Bm}{\Dstarm\Dz}$} \\
Double charm model& $0.01$ & $0.02$ & $0.11$ & $0.08$ & $0.21$ & $0.18$ \\
Single charm model& $0.05$ & $0.05$  & -  & -  & -  & - \\
Cross-feed model & -  & - & $0.00$ & $0.00$  & -  & - \\
Combinatorial model& $0.01$ & $0.00$ & $0.14$ & $0.13$ & $0.52$ & $0.15$ \\
External inputs& $0.05$ & $0.06$ & $0.01$ & $0.00$ & $0.04$ & $0.01$ \\
Total & $0.07$ & $0.08$ & $0.18$ & $0.15$ & $0.56$ & $0.23$ \\
\hline \vspace{-0.45cm} \\
& \multicolumn{2}{c}{$\decay{\Bm}{\Dsm\Dstarz}$} 
& \multicolumn{2}{c}{$\decay{\Bm}{\Dm\Dstarz}$} 
& \multicolumn{2}{c}{$\decay{\Bm}{\Dstarm\Dstarz}$} \\
Double charm model& $0.14$ & $0.23$ & $1.10$ & $1.04$ & $1.51$ & $1.19$ \\
Cross-feed model & -  & - & $0.00$ & $0.00$  & -  & - \\
Combinatorial model& $0.08$ & $0.04$ & $1.14$ & $0.92$ & $4.49$ & $1.02$ \\
External inputs& $0.55$ & $0.38$ & $0.14$ & $0.12$ & $0.11$ & $0.11$ \\
Total & $0.57$ & $0.45$ & $1.59$ & $1.40$ & $4.74$ & $1.57$ \\
\hline \vspace{-0.45cm} \\
& \multicolumn{2}{c}{$\decay{\Bm}{\Dssm\Dz}$} & & & &\\
Double charm model& $0.78$ & $0.51$ &&&&\\
Combinatorial model& $0.55$ & $0.22$ &&&&\\
External inputs& $0.89$ & $0.73$ &&&&\\
Total & $1.31$ & $0.91$ &&&&\\
\hline
\end{tabular}
\end{center}
\label{tab:systACPraw}
\end{table}

\begin{table}[tb]
    \caption{Systematic uncertainties on the corrections for \ACP in percent, averaged over all \Dz decay modes.}
    \begin{center}
\begin{tabular}{lcccccc}
\hline \vspace{-0.45cm} \\
\multirow{2}{*}{Final state} & \multicolumn{2}{c}{$\Dsm\Dz$} & \multicolumn{2}{c}{$\Dm\Dz$} & \multicolumn{2}{c}{$\Dstarm\Dz$}\\
                        & \multicolumn{1}{c}{Run~1} & \multicolumn{1}{c}{Run~2} & \multicolumn{1}{c}{Run~1} & \multicolumn{1}{c}{Run~2} & \multicolumn{1}{c}{Run~1} & \multicolumn{1}{c}{Run~2} \\
\hline
\AP& $0.42$ & $0.43$ & $0.41$ & $0.43$ & $0.48$ & $0.48$ \\
$\ACP(\decay{\Bp}{\jpsi\Kp})$& $0.30$ & $0.30$ & $0.30$ & $0.30$ & $0.30$ & $0.30$ \\
\AKpi& $0.28$ & $0.11$ & $0.04$ & $0.04$ & $0.10$ & $0.00$ \\
\Api& $0.09$ & $0.09$ & $0.06$ & $0.06$ & $0.18$ & $0.17$ \\
\APID& $0.29$ & $0.03$ & $0.25$ & $0.11$ & $0.55$ & $0.10$ \\
\ATIS& $0.08$ & $0.10$ & $0.08$ & $0.10$ & $0.09$ & $0.11$ \\
\ATOS& $0.01$ & $0.03$ & $0.01$ & $0.02$ & $0.01$ & $0.01$ \\
Weighting& $0.01$ & $0.00$ & $0.04$ & $0.00$ & $0.01$ & $0.00$ \\
Part. rec. weighting& $0.03$ & $0.02$ & $0.02$ & $0.01$ & $0.03$ & $0.01$ \\
Total & $0.67$ & $0.55$ & $0.58$ & $0.55$ & $0.82$ & $0.61$ \\

\hline
\end{tabular}
\end{center}
\label{tab:systACPcor}
\end{table}

\begin{table}[tb]
    \caption{Relative systematic uncertainties on branching fraction ratios, in percent, averaged over all \Dz decay modes.}
    \begin{center}
\begin{tabular}{lcccc}
\hline \vspace{-0.45cm} \\
\multirow{2}{*}{Source} & \multicolumn{2}{c}{$R(\Dm\Dz/\Dsm\Dz)$} & \multicolumn{2}{c}{$R(\Dstarm\Dz/\Dm\Dz)$} \\
& Run~1 & Run~2 & Run~1 & Run~2 \\
\hline
Double charm model& $0.8$ & $0.1$ & $1.2$ & $0.4$ \\
Single charm model& $0.3$ & $0.3$  & -  & - \\
Cross-feed model& $0.0$ & $0.0$ & $0.0$ & $0.0$ \\
Combinatorial model& $0.1$ & $0.0$ & $0.5$ & $0.2$ \\
Hardware trigger efficiency& $0.6$ & $0.1$ & $1.4$ & $0.2$ \\
Tracking efficiency& $0.4$ & $0.3$ & $1.4$ & $1.1$ \\
PID efficiency& $0.8$ & $0.6$ & $1.6$ & $0.7$ \\
Simulation sample size& $0.8$ & $0.9$ & $1.1$ & $1.1$ \\
Multiple candidate removal& $0.1$ & $0.1$ & $0.2$ & $0.4$ \\
BDT variable mismodelling& $0.8$ & $0.1$ & $1.0$ & $0.4$ \\
Track \chisq mismodelling& $0.5$ & $0.1$ & $0.5$ & $0.1$ \\
Weighting& $0.1$ & $0.0$ & $0.1$ & $0.0$ \\
\hline
Total & $1.8$ & $1.2$ & $3.3$ & $1.9$ \\
\hline
\end{tabular}
\end{center}
\label{tab:systBF}
\end{table}

The variation in the raw asymmetries and yields when using the alternative models for the invariant-mass distribution of signal and background contributions described below are assigned as the systematic uncertainties associated with the choice of invariant-mass shapes.
For fully reconstructed double-charm decays and double-charm decays with one missing particle, the resolution is described with a double Gaussian function instead of a Gaussian plus DSCB function.
For \decay{\B}{\Dstar\Dstar} decays with two missing particles and \decay{\B}{\Dstst\D} decays, the effect of detector resolution is removed from the invariant-mass distribution.
The systematic uncertainty associated with the incomplete knowledge of the branching fractions of \mbox{\decay{\B}{\Dstst\D}} decays~\cite{PDG2022} is estimated as the standard deviation of the asymmetries obtained when including only one \mbox{\decay{\B}{\Dstst\D}} decay at a time.
This results in the largest systematic uncertainty on \ACP from the double charm model.
The combinatorial background is described by a linear instead of an exponential function.

The yield and width of the single-charm background are varied by their statistical uncertainties.
The cross-feed is described by a single instead of a double Gaussian function and its yield is varied by its statistical uncertainty.
Systematic uncertainties are assigned for all external inputs that are fixed in the model as the change in raw asymmetry when varying the numerical values of the external inputs by their uncertainties.
These external inputs include asymmetries, branching fractions, polarisation fractions, $f_s/f_d$, the dilution factor and efficiency ratios.
For parameters where no external input is used, the assigned uncertainty covers the full range of possible values, namely 100\% for asymmetries, 50\% for polarisation fractions and 0.5 for the dilution factor.
The largest systematic uncertainties arise due to the absence of knowledge on \mbox{$\ACP(\decay{\Bs}{\Dssm\Dstarp})$} and \mbox{$f_L(\decay{\Bs}{\Dstarm\Dstarp})$}.

Statistical uncertainties from the size of the calibration and simulated signal samples used to obtain asymmetry and efficiency corrections are propagated as systematic uncertainties.
Additionally, each correction involves the measurement of the asymmetry in discrete phase space bins.
The finite size of these bins may introduce a bias, which is estimated and assigned as a systematic uncertainty by doubling the bin size in each dimension.
The size of the simulation samples results in negligible systematic uncertainties on the corrections to the \CP asymmetries.

There is an uncertainty of 0.3\% on $\AP(\Bm)$ from the uncertainty on \mbox{$\ACP(\decay{\Bp}{\jpsi\Kp})$}~\cite{PDG2022}.
The correction of $\Araw(\Kzb)$ to \AKpi carries uncertainties from detector material modelling and knowledge of the dependence of the asymmetry on the kinematics of the \Kzb, which are overall negligible.
A bias may arise from the assumption that the efficiency of PID requirements for all tracks are independent, when in reality, for example, the Cherenkov rings may overlap.
The size of this bias is measured in simulation and is found to be consistent with zero.
The uncertainty on the bias is assigned as a systematic uncertainty and is one of the largest uncertainties on the PID asymmetries and efficiencies.
The use of PID requirements that double the inefficiency compared to the effective PID requirements results in a change of up to 0.1\% in \APID, and this difference is assigned as the systematic uncertainty corresponding to the definition of the effective PID requirement.
Not applying the BDT selection when determining \APID softens the kinematic spectrum, therefore, a systematic uncertainty is assigned as the difference when the BDT requirement is applied, which is found to be negligible.

The efficiency of the hadron hardware trigger depends on the transverse energy and position at the surface of the HCAL of all final-state particles in a decay.
Use of the simulated TIS candidates to evaluate \ATOS and the efficiencies of the TOS and xTIS requirements necessitates the assumption that this dependence can be simplified to a variation in bins of \mbox{$(\pt(\Bm),\eta(\Bm))$}.
The bias associated with this simplification is estimated as the change when the measurement is instead binned in the maximum transverse energy of the final-state particles that are within the HCAL acceptance.
Corrections for overlapping energy deposits and HCAL cell miscalibration are not included in the evaluation of the hardware trigger TOS requirement efficiency for Run~1 data, and the resulting bias is estimated as the effect of not including these corrections for Run~2 data.
These two contributions are the dominant sources of systematic uncertainty on the hardware trigger efficiency and are negligible when considering \ATOS.

The uncertainty on the material budget results in a 1.4\% (1.1\%) uncertainty on the fraction of kaons (pions) that are not reconstructed because they undergo a hadronic interaction before the final tracking station.
This results in an uncertainty of the same size per final-state particle on the efficiencies.
Correcting the calibration simulation in different event multiplicity variables results in small variations in the tracking efficiency~\cite{LHCb-DP-2013-002}, which are propagated as a systematic uncertainty.

The efficiency of removing multiple candidates is not well modelled by simulation, so the fraction of candidates removed is assigned as a systematic uncertainty.
The value of \chisqip of the \Bm candidate, and the vertex fit quality and $t/\sigma_t$ of the \Dz candidate are used as inputs to the BDT classifiers and their distributions in simulation do not match those in data.
A systematic uncertainty is attributed as the change in efficiency when the simulation is corrected as a function of these distributions with a weighting procedure.
The \chisq quality of the track fit is also poorly reproduced, and a requirement is applied on this variable in the offline selection.
The inefficiency of this requirement is estimated in background-subtracted data and simulation by extrapolating the distribution.
The inefficiencies in data and simulation are propagated to an estimate of the bias on the simulated efficiency that is assigned as a systematic uncertainty.

The uncertainty on the weights to correct the production distribution and four-body \Dz decay are determined by resampling the training dataset~\cite{efron:1979} and changing the model in the background subtraction.
For partially reconstructed decays, the systematic uncertainty on the weight function that relates the kinematics of fully and partially reconstructed decays is established by varying the power-law parameter by its statistical uncertainty and by performing the correction in $\pt(\Bm)$ instead of $p(\Bm)$.
A correction in the kinematics of the \Bm meson cannot resolve discrepancies in the kinematics of the final-state particles, which is relevant to detection asymmetries depending on the final-state particle kinematic distributions.
However, this correction results in a change to the final-state particle kinematic distributions of similar size to the discrepancies.
Therefore, for the relevant detection asymmetries, the difference with and without the weights for partially reconstructed decay kinematics is assigned as a systematic uncertainty.
In all cases, the systematic uncertainty associated with both weighting procedures is negligible.

\section{Results and conclusions}

The \CP asymmetries are measured by correcting the raw asymmetries for the production and detection asymmetries using Eq.~\ref{eq:ACP}.
Measurements from different \Dz decay channels and different data taking periods are combined taking into account correlations of the systematic uncertainties~\cite{Valassi:2014}.
The resulting measurements are
\begin{equation*}
\begin{split}
    \ACP(\decay{\Bm}{\Dsm\Dz}) & = (+0.5\pm0.2\pm0.5\pm0.3)\%, \\
    \ACP(\decay{\Bm}{\Dssm\Dz}) & = (-0.5\pm1.1\pm1.0\pm0.3)\%, \\
    \ACP(\decay{\Bm}{\Dsm\Dstarz}) & = (+1.1\pm0.8\pm0.6\pm0.3)\%, \\
    \ACP(\decay{\Bm}{\Dm\Dz}) & = (+2.5\pm1.0\pm0.4\pm0.3)\% ,\\
    \ACP(\decay{\Bm}{\Dm\Dstarz}) & = (-0.2\pm2.0\pm1.4\pm0.3)\%, \\
    \ACP(\decay{\Bm}{\Dstarm\Dz}) & = (+3.3\pm1.6\pm0.6\pm0.3)\% ,\\
    \ACP(\decay{\Bm}{\Dstarm\Dstarz}) & = (+2.3\pm2.1\pm1.7\pm0.3)\%, \\
    \end{split}
\end{equation*}
where the first uncertainty is statistical, the second is systematic, and the third is from \mbox{$\ACP(\decay{\Bp}{\jpsi\Kp})$}.
The associated correlations are given in Table~\ref{tab:corr}.
No evidence of \CP violation is seen.
The \ACP for each decay is compared between Run~1 and Run~2, final states and magnet polarities, and no difference is larger than two standard deviations.

\begin{table}[tb]
\centering
\caption{Total correlations between the measured \ACP.}
\label{tab:corr}
\begin{tabular}{l|ccccccc}
    \hline
    & $\Dsm\Dz$ & $\Dssm\Dz$ & $\Dsm\Dstarz$ & $\Dm\Dz$ & $\Dm\Dstarz$ & $\Dstarm\Dz$ & $\Dstarm\Dstarz$ \\
    \hline
$\Dsm\Dz$ & $\phantom{-}1\phantom{.000}$ & & & & & & \\
$\Dssm\Dz$ & $\phantom{-}0.335$ & $\phantom{-}1\phantom{.000}$ & & & & & \\
$\Dsm\Dstarz$ & $\phantom{-}0.431$ & $-0.282$ & $\phantom{-}1\phantom{.000}$ & & & & \\
$\Dm\Dz$ & $\phantom{-}0.386$ & $\phantom{-}0.186$ & $\phantom{-}0.234$ & $\phantom{-}1\phantom{.000}$ & & & \\
$\Dm\Dstarz$ & $\phantom{-}0.183$ & $\phantom{-}0.227$ & $\phantom{-}0.180$ & $\phantom{-}0.195$ & $\phantom{-}1\phantom{.000}$ & & \\
$\Dstarm\Dz$ & $\phantom{-}0.286$ & $\phantom{-}0.161$ & $\phantom{-}0.177$ & $\phantom{-}0.166$ & $\phantom{-}0.130$ & $\phantom{-}1\phantom{.000}$ & \\
$\Dstarm\Dstarz$ & $\phantom{-}0.190$ & $\phantom{-}0.227$ & $\phantom{-}0.211$ & $\phantom{-}0.168$ & $\phantom{-}0.311$ & $\phantom{-}0.189$ & $\phantom{-}1\phantom{.000}$ \\
    \hline
\end{tabular}
\end{table}

The measurement of \mbox{$R(\Dm\Dz/\Dsm\Dz)$} is presented in Table~\ref{tab:RDmDz} and \mbox{$R(\Dstarm\Dz/\Dm\Dz)$} in Table~\ref{tab:RDstmDz}.
These results are in agreement with the world averages.
The correlation between the measured \mbox{$R(\Dm\Dz/\Dsm\Dz)$} and \mbox{$R(\Dstarm\Dz/\Dm\Dz)$} is $-47\%$.
Measurements of $R(\Dm\Dz/\Dsm\Dz)$ and $R(\Dstarm\Dz/\Dm\Dz)$ are compared between Run~1 and Run~2 data, different \Dz decay channels and different hardware trigger decisions.
All measurements are consistent within three standard deviations.
The effect upon the branching fraction ratios of varying the BDT and PID requirements is also studied.
In total, 48 tests are performed, of which none results in a change in the central value of more than three standard deviations.

\begin{table}[tb]
    \centering
    \caption{$R(\Dm\Dz/\Dsm\Dz)/10^{-2}$ for each \Dz decay mode, for Run~1 and Run~2 and the combined measurement.
The first uncertainty is statistical and the second is systematic.}
\label{tab:RDmDz}
    \begin{tabular}{lccc}
        \hline
        \Dz decay mode & Run~1 & Run~2 & Run~1+2 \\
        \hline
$\Km\pip$ & $ 6.88\pm0.24\pm0.12$ & $ 7.35\pm0.12\pm0.11$ & $ 7.22\pm0.11\pm0.10$ \\
$\Km\pip\pim\pip$ & $ 6.93\pm0.38\pm0.23$ & $ 7.40\pm0.18\pm0.15$ & $ 7.30\pm0.16\pm0.14$ \\
Combined & $ 6.89\pm0.20\pm0.12$ & $ 7.36\pm0.10\pm0.10$ & $ 7.25\pm0.09\pm0.09$ \\

        \hline
    \end{tabular}
\end{table}

\begin{table}[tb]
    \centering
    \caption{$R(\Dstarm\Dz/\Dm\Dz)$ for each \Dz decay mode, for Run~1 and Run~2 and the combined measurement.
The first uncertainty is statistical and the second is systematic.}
\label{tab:RDstmDz}
    \begin{tabular}{lccc}
        \hline
        \Dz decay mode & Run~1 & Run~2 & Run~1+2 \\
        \hline
        $\Km\pip$ & $ 0.328\pm0.023\pm0.011$ & $ 0.256\pm0.009\pm0.005$ & $ 0.271\pm0.008\pm0.005$ \\
$\Km\pip\pim\pip$ & $ 0.316\pm0.033\pm0.015$ & $ 0.272\pm0.012\pm0.008$ & $ 0.278\pm0.012\pm0.007$ \\
Combined & $ 0.324\pm0.019\pm0.010$ & $ 0.262\pm0.007\pm0.005$ & $ 0.271\pm0.007\pm0.005$ \\
        \hline
    \end{tabular}
\end{table}

In conclusion, the most precise \CP asymmetry measurements for seven \mbox{\decay{\Bm}{\DmorDsorDstar\DzorDstar}} decays are performed, including the first measurements of \mbox{$\ACP(\decay{\Bm}{\Dssm\Dz})$} and \mbox{$\ACP(\decay{\Bm}{\Dsm\Dstarz})$}.
No evidence for \CP violation is found.
The measurements of \mbox{$\ACP(\decay{\Bm}{\DmorDs\Dz})$} are in agreement with and supersede the earlier \lhcb measurement that uses only Run~1 data~\cite{LHCb-PAPER-2018-007}.
Ratios between the branching fractions of fully reconstructed \decay{\Bm}{\DmorDsorDstar\Dz} decays are also measured.
The measurements presented in this paper substantially improve knowledge of \Bm meson decays to two charm mesons, which can help to constrain BSM models~\cite{Xu:2016hpp,Kim:2008ex,Lu:2010gg}.

\section*{Acknowledgements}
%
%
\noindent We express our gratitude to our colleagues in the CERN
accelerator departments for the excellent performance of the LHC. We
thank the technical and administrative staff at the LHCb
institutes.
We acknowledge support from CERN and from the national agencies:
CAPES, CNPq, FAPERJ and FINEP (Brazil); 
MOST and NSFC (China); 
CNRS/IN2P3 (France); 
BMBF, DFG and MPG (Germany); 
INFN (Italy); 
NWO (Netherlands); 
MNiSW and NCN (Poland); 
MEN/IFA (Romania); 
MICINN (Spain); 
SNSF and SER (Switzerland); 
NASU (Ukraine); 
STFC (United Kingdom); 
DOE NP and NSF (USA).
We acknowledge the computing resources that are provided by CERN, IN2P3
(France), KIT and DESY (Germany), INFN (Italy), SURF (Netherlands),
PIC (Spain), GridPP (United Kingdom), 
CSCS (Switzerland), IFIN-HH (Romania), CBPF (Brazil),
Polish WLCG  (Poland) and NERSC (USA).
We are indebted to the communities behind the multiple open-source
software packages on which we depend.
Individual groups or members have received support from
ARC and ARDC (Australia);
Minciencias (Colombia);
AvH Foundation (Germany);
EPLANET, Marie Sk\l{}odowska-Curie Actions, ERC and NextGenerationEU (European Union);
A*MIDEX, ANR, IPhU and Labex P2IO, and R\'{e}gion Auvergne-Rh\^{o}ne-Alpes (France);
Key Research Program of Frontier Sciences of CAS, CAS PIFI, CAS CCEPP, 
Fundamental Research Funds for the Central Universities, 
and Sci. \& Tech. Program of Guangzhou (China);
GVA, XuntaGal, GENCAT, Inditex, InTalent and Prog.~Atracci\'on Talento, CM (Spain);
SRC (Sweden);
the Leverhulme Trust, the Royal Society
 and UKRI (United Kingdom).


\addcontentsline{toc}{section}{References}
\bibliographystyle{LHCb}
\bibliography{main,standard,LHCb-PAPER,LHCb-CONF,LHCb-DP,LHCb-TDR}

\newpage
\centerline
{\large\bf LHCb collaboration}
\begin
{flushleft}
\small
R.~Aaij$^{32}$\lhcborcid{0000-0003-0533-1952},
A.S.W.~Abdelmotteleb$^{51}$\lhcborcid{0000-0001-7905-0542},
C.~Abellan~Beteta$^{45}$,
F.~Abudin{\'e}n$^{51}$\lhcborcid{0000-0002-6737-3528},
T.~Ackernley$^{55}$\lhcborcid{0000-0002-5951-3498},
B.~Adeva$^{41}$\lhcborcid{0000-0001-9756-3712},
M.~Adinolfi$^{49}$\lhcborcid{0000-0002-1326-1264},
P.~Adlarson$^{78}$\lhcborcid{0000-0001-6280-3851},
H.~Afsharnia$^{9}$,
C.~Agapopoulou$^{43}$\lhcborcid{0000-0002-2368-0147},
C.A.~Aidala$^{79}$\lhcborcid{0000-0001-9540-4988}
% ...

{\footnotesize \it

% ...
$^{9}$Universit{\'e} Clermont Auvergne, CNRS/IN2P3, LPC, Clermont-Ferrand, France\\
% ...
$^{32}$Nikhef National Institute for Subatomic Physics, Amsterdam, Netherlands\\
% ...
$^{41}$Instituto Galego de F{\'\i}sica de Altas Enerx{\'\i}as (IGFAE), Universidade de Santiago de Compostela, Santiago de Compostela, Spain\\
% ...
$^{43}$European Organization for Nuclear Research (CERN), Geneva, Switzerland\\
% ...
$^{45}$Physik-Institut, Universit{\"a}t Z{\"u}rich, Z{\"u}rich, Switzerland\\
% ...
$^{49}$H.H. Wills Physics Laboratory, University of Bristol, Bristol, United Kingdom\\
% ...
$^{51}$Department of Physics, University of Warwick, Coventry, United Kingdom\\
% ...
$^{54}$School of Physics and Astronomy, University of Glasgow, Glasgow, United Kingdom\\
$^{55}$Oliver Lodge Laboratory, University of Liverpool, Liverpool, United Kingdom\\
% ...
$^{63}$Syracuse University, Syracuse, NY, United States\\
% ...
$^{78}$Department of Physics and Astronomy, Uppsala University, Uppsala, Sweden, associated to $^{54}$\\
$^{79}$University of Michigan, Ann Arbor, MI, United States, associated to $^{63}$\\
\bigskip
% ...
}
\end{flushleft}

\end{document}